\documentclass[prd,nofootinbib,twocolumn]{revtex4}
\usepackage{graphicx}
\usepackage{bm}
\usepackage{float}
\usepackage{subfigure}
\usepackage{amsmath}
\usepackage{amsfonts}
\usepackage{color}
\usepackage{slashed}
\usepackage{fancyhdr}

\newcommand{\gb}{\bar{g}}

\newcommand{\dd}{\text{d}}

\newcommand{\nn}{\nonumber}
\newcommand{\nnn}{\nonumber \\}

\newcommand{\bsigma}{\bar{\sigma}}
\newcommand{\bG}{\bar{G}}
\newcommand{\Db}{\overline{\nabla}}

\begin{document}

\title{Retarded Green's Functions In Perturbed Spacetimes For Cosmology and Gravitational Physics}

\author{Yi-Zen Chu$^{1,2}$ and Glenn D. Starkman$^\dagger$}
\affiliation{
$^1$Center for Particle Cosmology, Department of Physics and Astronomy, University of Pennsylvania, Philadelphia, Pennsylvania 19104, USA, \\
$^2$Physics Department, Arizona State University, Tempe, AZ 85287, USA \\
$^\dagger$CERCA, Physics Department, Case Western Reserve University, Cleveland, OH 44106-7079, USA
}

\begin{abstract}
\noindent Electromagnetic and gravitational radiation do not propagate solely on the null cone in a generic curved spacetime. They develop ``tails," traveling at all speeds equal to and less than unity. If sizeable, this off-the-null-cone effect could mean objects at cosmological distances, such as supernovae, appear dimmer than they really are. Their light curves may be distorted relative to their flat spacetime counterparts. These in turn could affect how we infer the properties and evolution of the universe or the objects it contains. Within the gravitational context, the tail effect induces a self-force that causes a compact object orbiting a massive black hole to deviate from an otherwise geodesic path. This needs to be taken into account when modeling the gravitational waves expected from such sources. Motivated by these considerations, we develop perturbation theory for solving the massless scalar, photon and graviton retarded Green's functions in perturbed spacetimes $g_{\mu\nu} = \gb_{\mu\nu} + h_{\mu\nu}$, assuming these Green's functions are known in the background spacetime $\gb_{\mu\nu}$. In particular, we elaborate on the theory in perturbed Minkowski spacetime in significant detail; and apply our techniques to compute the retarded Green's functions in the weak field limit of the Kerr spacetime to first order in the black hole's mass $M$ and angular momentum $S$. Our methods build on and generalizes work appearing in the literature on this topic to date, and lays the foundation for a thorough, first principles based, investigation of how light propagates over cosmological distances, within a spatially flat inhomogeneous Friedmann-Lema\^{i}tre-Robertson-Walker (FLRW) universe. This perturbative scheme applied to the graviton Green's function, when pushed to higher orders, may provide approximate analytic (or semi-analytic) results for the self-force problem in the weak field limits of the Schwarzschild and Kerr black hole geometries.
\end{abstract}

\maketitle

\section{Introduction and Motivation}

This paper is primarily concerned with how to solve for the retarded Green's functions of the minimally coupled massless scalar $\varphi$, photon $A_\mu$, and graviton $\gamma_{\mu\nu}$ in spacetimes described by the perturbed metric $g_{\mu\nu} = \gb_{\mu\nu} + h_{\mu\nu}$, if the solutions are known in the background metric $\gb_{\mu\nu}$. One important instance is that of Minkowski spacetime, where these Green's functions\footnote{Since we will be dealing exclusively with retarded Green's functions, we will drop the word ``retarded" from henceforth. Despite this restriction, our methods actually apply for advanced Green's functions too.} are known explicitly. Here, we will carry out the analysis in detail for the 4 dimensional case, and obtain $\mathcal{O}[h]$-accurate solutions to the Green's functions in perturbed Minkowski spacetime up to quadrature. Our methods are akin to the Born approximation employed in quantum theory, where one first obtains an integral equation for the Green's functions, and the $\mathcal{O}[h^N]$-accurate answer is gotten after $N$ iterations, followed by dropping a remainder term. We are not the first to develop perturbation theory for solving Green's functions about weakly curved spacetimes. DeWitt and DeWitt \cite{DeWittDeWitt:1964}, Kovacs and Thorne \cite{KovacsThorne:1975}, and more recently, Pfenning and Poisson \cite{PfenningPoisson:2000zf} have all tackled this problem using various techniques which we will briefly compare against in the conclusions. As far as we are aware, however, our approach is distinct from theirs and have not appeared before in the gravitational physics and cosmology literature.

Green's functions play crucial roles in understanding the dynamics of both classical and quantum field theories. The Green's function depends on the coordinates of two spacetime locations we will denote as $x \equiv (t,\vec{x})$ and $x' \equiv (t',\vec{x}')$,\footnote{The spacetime coordinates in this paper will take the form $x$, $x'$, $x''$, etc. Instead of displaying the dependence on these coordinates explicitly, we will put primes on the indices of tensorial quantities to indicate which of the variables are to be associated with them. For example $G_{\mu\nu'} = G_{\mu\nu}[x,x']$, $\nabla_{\mu''}$ denotes the covariant derivative with respect to $x''$, $g$ is the determinant of the metric at $x$ and $g'$ at $x'$, etc.} and respectively identify as the observer and source positions. At the classical level, which will be the focus of this paper, it can be viewed as the field measured at the spacetime point $x$ produced by a spacetime-point source with unit charge at $x'$. To understand this, consider some spacetime region $V$ between two constant time hypersurfaces $t$ and $t'$, with $t > t'$. In this paper we assume that spacetime is an infinite (or, in the cosmological context, semi-infinite) manifold. Let there be some field producing source $J$ present in the volume $V$, and non-trivial initial conditions for the fields at $t'$, for example, $\varphi[x'^0 = t']$ and $\nabla^{0'} \varphi[x'^0 = t']$. Denote the scalar, photon, and graviton Green's functions as $G_{x,x'}$, $G_{\mu\nu'}$ and $G_{\delta\epsilon \rho'\sigma'}$ respectively. Then the scalar field $\varphi$ evaluated at some point $x$ lying on the $t$ surface can be written as\footnote{These are known as the Kirchhoff representations. We refer the reader to the review by Poisson \cite{PoissonReview:2003nc} for their derivation. In this paper, whenever a formula holds in arbitrary spacetime dimensions greater or equal to 4, we will use $d$ to denote the dimensions of spacetime. Summation convention is in force: Greek letters run from 0 to $d-1$ while small English alphabets run from 1 to $d-1$.}
{\allowdisplaybreaks\begin{align}
\label{KirchoffRep_Scalar}
&\varphi_x = \int_V \dd^d x' |g'|^{\frac{1}{2}} G_{x,x'} J_{x'} \\
&+ \left.\int \dd^{d-1} \vec{x}' |g'|^{\frac{1}{2}}
    \left( G_{x,x'} \nabla^{0'} \varphi_{x'} - \nabla^{0'} G_{x,x'} \varphi_{x'} \right)\right\vert_{x'^0 = t'} \nn
\end{align}}
while the photon's vector potential evaluated at $x$ can be written as
{\allowdisplaybreaks\begin{align}
\label{KirchoffRep_Photon}
&A_\mu = \int_V \dd^d x' |g'|^{\frac{1}{2}} G_{\mu \nu'} J^{\nu'} \\
&+ \left.\int \dd^{d-1} \vec{x}' |g'|^{\frac{1}{2}}
    \left( G_{\mu \nu'} \nabla^{0'} A^{\nu'} - \nabla^{0'} G_{\mu\nu'} A^{\nu'}\right)\right\vert_{x'^0 = t'} . \nn
\end{align}}
In the same vein, the graviton field at $x$ reads\footnote{We will not be concerned with the nonlinear self-interaction of the gravitons in this paper. However, these nonlinear terms may be considered to be part of $J_{\mu\nu}$, since gravity gravitates.}
{\allowdisplaybreaks\begin{align}
\label{KirchoffRep_Graviton}
\gamma_{\mu\nu}
&= \int_V \dd^d x' |g'|^{\frac{1}{2}} G_{\mu\nu \alpha'\beta'} J^{\alpha'\beta'} \\
&+ \int \dd^{d-1} \vec{x}' |g'|^{\frac{1}{2}}
    \bigg( G_{\mu\nu \alpha'\beta'} P^{\alpha'\beta'}_{\phantom{\alpha'\beta'}\rho'\epsilon'} \nabla^{0'} \gamma^{\rho'\epsilon'} \nnn
    &\qquad \qquad
        - \nabla^{0'} G_{\mu\nu \alpha'\beta'} P^{\alpha'\beta'}_{\phantom{\alpha'\beta'}\rho'\epsilon'} \gamma^{\rho'\epsilon'} \left.\bigg)\right\vert_{x'^0 = t'} , \nn
\end{align}}
where $P^{\alpha'\beta'}_{\phantom{\alpha'\beta'}\rho'\epsilon'} \equiv (1/2)(\delta^{\alpha}_{\rho}\delta^{\beta}_{\epsilon} + \delta^{\alpha}_{\epsilon}\delta^{\beta}_{\rho} - g^{\alpha'\beta'}g_{\rho'\epsilon'})$.

\begin{figure}
\includegraphics[width=3.5in]{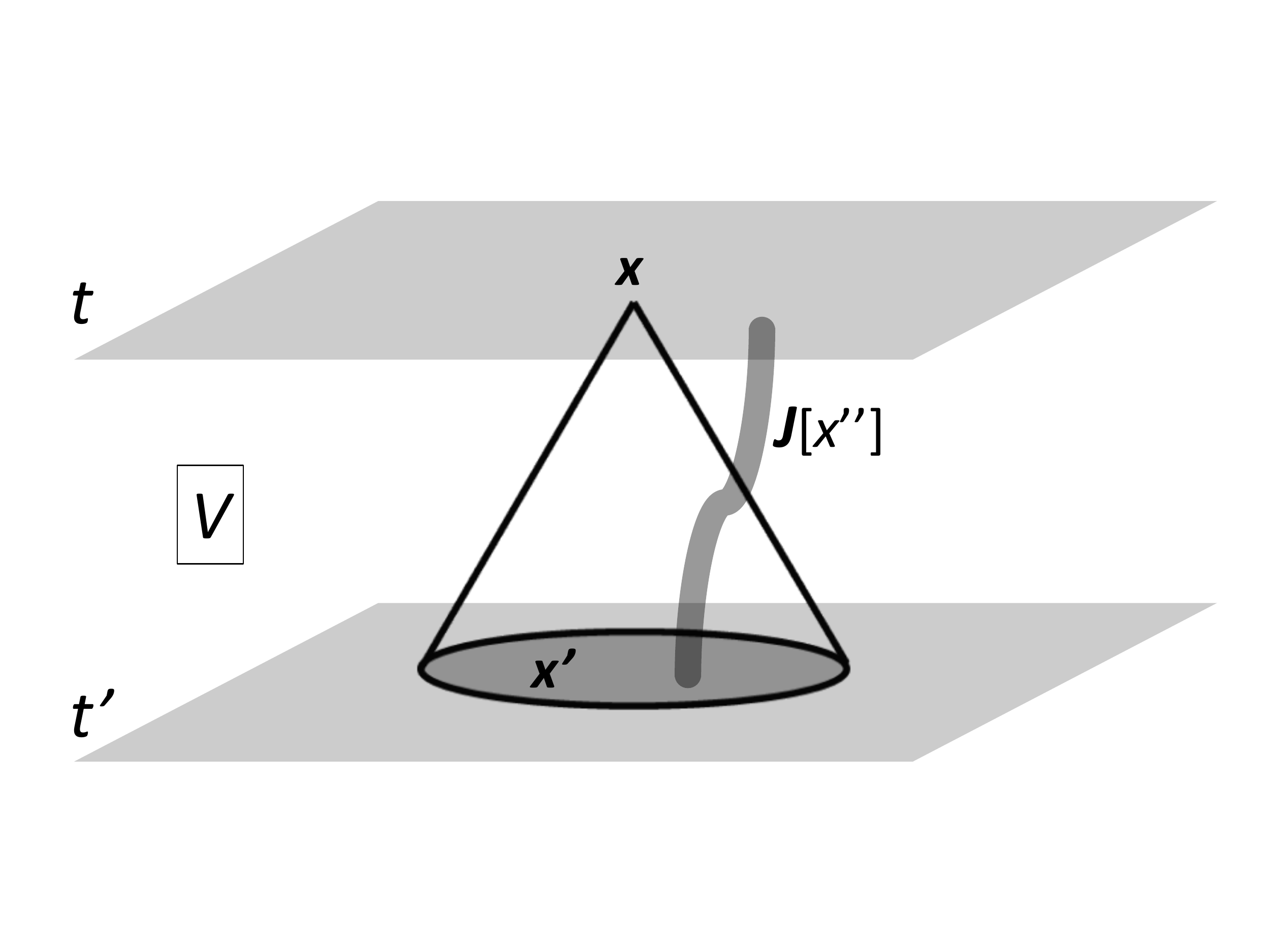}
\caption{The spacetime region $V$ is the volume contained within the two constant time surfaces at times $t$ and $t'$, where $t > t'$. The observer is located at $x \equiv (t,\vec{x})$. The dark oval is the region defined by the intersection between the past light cone of $x$ and its interior with that of the constant time surface at $t'$. On it, we allow some non-trivial field configuration to be present, and through the Kirchhoff representations in equations \eqref{KirchoffRep_Scalar}, \eqref{KirchoffRep_Photon} and \eqref{KirchoffRep_Graviton}, the Green's functions evolves it forward in time. We emphasize that the causal structure of the Green's function, as exhibited by equations \eqref{HadamardForm_Scalar}, \eqref{HadamardForm_Photon} and \eqref{HadamardForm_Graviton} means, in a generic curved spacetime, the observer receives fields not only from her past light cone (edge of the dark oval), but also its interior (dark oval itself). In addition, there is some (scalar, photon or graviton)-producing source $J$ which sweeps out a world tube, and our observer receives radiation from the portion of this world tube that lies on and within the interior of her past light cone. The picture here is to be contrasted against the Minkowski one, where observers only detect fields from their past null cone.}
\label{InitialValueFigure}
\end{figure}
From equations \eqref{KirchoffRep_Scalar}, \eqref{KirchoffRep_Photon} and \eqref{KirchoffRep_Graviton}, we see that the physical solution of a linear field theory can be expressed as the sum of two integrals of the Green's function (and its gradient), one weighted by the sources present in the system at hand and the other weighted by the initial conditions of the fields themselves. In particular, the $d$-dimensional volume integrals (with respect to $x'$) of the Green's functions, weighted by the field-producing $J$s, reaffirms the interpretation that the Green's function is the field of a spacetime-point ``unit charge" because these volume integrals corresponds to calculating the field at $x$ by superposing the field produced by all the ``charges" $J$s present in the system.

Moreover, that the Green's function yields a causality respecting solution can be seen from the following. In a generic curved spacetime, if $\sigma_{x,x'}$ (usually known as Sygne's world function) denotes half the square of the geodesic distance between $x$ and $x'$, a general analysis in 4 dimensions tells us that the Green's function in a generic curved spacetime consists of two terms.\footnote{See Poisson's review \cite{PoissonReview:2003nc} for the Hadamard construction of the Green's functions in \eqref{HadamardForm_Scalar}, \eqref{HadamardForm_Photon} and \eqref{HadamardForm_Graviton} below. We note in passing that, in higher than 4 dimensions, the general form of the Green's function will be more complicated, containing not only $\delta$ and $\Theta$ terms, but derivatives of $\delta$-functions too.} One of them is proportional to $\Theta[t-t'] \delta[\sigma_{x,x'}]$, and describes propagation of the fields on the null cone. The other is proportional to $\Theta[t-t'] \Theta[\sigma_{x,x'}]$, and describes propagation on the interior of the future light cone of $x'$.
{\allowdisplaybreaks\begin{align}
\label{HadamardForm_Scalar}
G_{x,x'} &= \frac{\Theta[t-t']}{4\pi} \left( U_{x,x'} \delta[\sigma_{x,x'}] + V_{x,x'} \Theta[\sigma_{x,x'}] \right), \\
\label{HadamardForm_Photon}
G_{\mu\nu'} &= \frac{\Theta[t-t']}{4\pi} \left( U_{\mu\nu'} \delta[\sigma_{x,x'}] + V_{\mu\nu'} \Theta[\sigma_{x,x'}] \right), \\
\label{HadamardForm_Graviton}
G_{\mu\nu \alpha'\beta'} &= \frac{\Theta[t-t']}{4\pi} \big( U_{\mu\nu \alpha'\beta'} \delta[\sigma_{x,x'}] \nnn
&\qquad \qquad \qquad \qquad + V_{\mu\nu \alpha'\beta'} \Theta[\sigma_{x,x'}] \big).
\end{align}}
In Fig. \eqref{InitialValueFigure}, we illustrate the Kirchhoff representations in \eqref{KirchoffRep_Scalar}, \eqref{KirchoffRep_Photon} and \eqref{KirchoffRep_Graviton}.

{\bf Tails in curved spacetime} \quad The presence of the two terms in equations \eqref{HadamardForm_Scalar}, \eqref{HadamardForm_Photon} and \eqref{HadamardForm_Graviton}, the property that for some fixed $x'$, the Green's functions of massless fields are non-zero for all $x$ both on and inside the future null cone for $x'$, teaches us an important difference between the propagation of electromagnetic and gravitational wave signals in a curved versus flat 4 dimensional spacetime. In the latter, signals travel strictly on the null cone, and the radiation received at some location $x$ is related to the source at retarded time $t-t' = |\vec{x}-\vec{x}'|$. In the former, signals travel at all speeds equal and less than unity.\footnote{This is barring special properties, such as the conformal symmetry enjoyed by the Maxwell action, which says that light is blind to conformal factors of the metric: $a^2 g_{\mu\nu}$ and $g_{\mu\nu}$ are equivalent in its eyes. We will shortly elaborate on this point.} (We are setting $c=1$.) This off-the-light-cone piece of massless radiation is known in the literature as the tail (or, sometimes, wake); and is often touted as a violation of Huygens' principle in curved spacetime. As elucidated by DeWitt and Brehme \cite{DeWittBrehme:1960}, this implies the electrodynamics of even a single electrically charged particle depends on its entire past history: it exerts a force upon itself (a ``self-force"), in addition to the one already present in flat Minkowski spacetime, because the electromagnetic fields it produces travels away from it but then scatters off the geometry of spacetime and returns to interact with it at some later time.

{\bf Gravitational Dynamics} \quad This tail-induced self-force finds an analog in the gravitational dynamics of compact objects orbiting massive black holes, because the gravitational waves they generate scatter off the non-trivial background geometry and return to nudge their trajectories away from a geodesic one. The gravitational radiation signals of such systems are believed to be within reach of future gravitational wave detectors, and there is currently intense theoretical work done to understand their dynamics. Perturbation theory in the weak field limit of Schwarzschild and Kerr may thus provide us with approximate but concrete results from which we can gain physical insight from (and possibly serve as a check against numerical calculations). For instance, that the tail effect is the result of massless fields scattering off the background geometry will be manifest within the perturbative framework we are about to undertake; this point has already been noted by DeWitt and DeWitt \cite{DeWittDeWitt:1964} and Pfenning and Poisson \cite{PfenningPoisson:2000zf}.

{\bf Cosmology} \quad Turning our attention now to cosmology, the past decades have provided us with observational evidence that we live in a universe that is, at the roughest level, described by the spatially flat FLRW metric. In conformal coordinates, it is
\begin{align}
\label{FLRWMetric}
g_{\mu\nu} = a^2 \eta_{\mu\nu}, \qquad \eta_{\mu\nu} \equiv \text{diag}[1,-1,-1,-1],
\end{align}
where $a$ tells us the relative size of the universe at various times along its evolution. Most of our inference of the properties of the universe come from examining light emanating from objects at cosmological or astrophysical distances, and furthermore our interpretation of electromagnetic signals are based on the assumption that they travel on null geodesics. This statement is precisely true when the metric is \eqref{FLRWMetric} because the Maxwell action that governs the dynamics of photons in vacuum, is insensitive to the conformal factor $a^2$. Specifically, in 4 dimensional spacetime, $S_\text{Maxwell}[\eta]$ and $S_\text{Maxwell}[a^2 \eta]$ are exactly the same object; the conformal factor $a^2$ drops out.
\begin{align}
\label{MaxwellAction_ConformalSymmetry}
&S^{(d=4)}_\text{Maxwell}[a^2 \eta] = S^{(d=4)}_\text{Maxwell}[\eta] \\
&= -\frac{1}{4} \int \dd^4 x'' \eta^{\mu''\alpha''} \eta^{\nu''\beta''} F_{\mu''\nu''} F_{\alpha''\beta''} \nn
\end{align}
This means electromagnetic radiation in 4 dimensional spatially flat FLRW universes behaves no differently from how it does in 4 dimensional Minkowski spacetimes. In particular, it travels only along null geodesics. However, cosmological and astrophysical observations have become so sensitive that it is no longer sufficient to model our universe as the exactly smooth and homogeneous spacetime in \eqref{FLRWMetric}. Rather, one needs to account for the metric perturbations,
\begin{align}
\label{FLRWMetric_Perturbed}
g_{\mu\nu} = a^2 \left( \eta_{\mu\nu} + h_{\mu\nu} \right).
\end{align}
Because the $a^2$ drops out of the Maxwell action, we recognize that a first principles theoretical investigation of the propagation of light over cosmological distances is equivalent to the same investigation in perturbed Minkowski spacetime. Moreover, since the geometry is now curved (albeit weakly so), light traveling over cosmological length scales should therefore develop tails. As already mentioned in the abstract, if a significant portion of light emitted from a supernova at cosmological distances leaks off the light cone, then the observer on Earth may mistakenly infer that it is dimmer than it actually is, as some of the light has not yet arrived. This leakage may also modify the light curves of these objects at cosmological distances. To our knowledge, the size of the electromagnetic tail effect in cosmology has not been examined before. Our development of perturbation theory for the photon Green's function (and confirmation of DeWitt and DeWitt's first order results \cite{DeWittDeWitt:1964}) in perturbed Minkowski, is therefore the first step to a thorough, first principles based, understanding of the properties of light in the cosmological context. This may in turn affect how we interpret cosmological and astronomical observations.\footnote{We emphasize here that, we are not, as yet, claiming that the tail effect is significant in the cosmological context. Rather, this paper is laying down the groundwork -- the computation of the photon Green's function -- in order to investigate this issue from first principles.}

{\bf JWKB} \quad Now, the JWKB approximation where one assumes that the wavelength of the massless fields are extremely small relative to the characteristic length scales of the spacetime geometry (which, in term, usually amounts to neglecting all geometric terms relative to the $\Box$ in the wave equation), is often used to justify that null cone propagation is the dominant channel of travel for massless fields in generic curved spacetimes. (See for example Misner, Thorne and Wheeler \cite{MisnerThorneWheeler:1974qy}.) Here, we caution that, even in cases where the JWKB approximation yields exact results, it does not imply that light travels solely on the light cone. Such a counterexample is that of odd dimensional Minkowski spacetimes, where the momentum vector $k_\mu$ satisfies the exact dispersion relation $\eta^{\mu\nu} k_\mu k_\nu = 0$, but the Green's functions of massless fields develop power law tails: for odd $d$, $V[x,x'] \propto ((t-t')^2-(\vec{x}-\vec{x}')^2)^{-(d-2)/2}$. (See Soodak and Tiersten \cite{SoodakTiersten} for a pedagogical discussion on tails of Green's functions in Minkowski spacetimes.) This tells us that, even for 4 dimensional flat spacetime, the rigorous way to prove that light travels on the null cone is by computing the photon Green's function, since it is the Green's function (via the Kirchhoff representations in \eqref{KirchoffRep_Scalar}, \eqref{KirchoffRep_Photon} and \eqref{KirchoffRep_Graviton}) that determines how physical signals propagate away from their sources.

In the next section, we will review the general theory of Green's functions and some geometrical constructs related to them. Perturbation theory for Green's functions will then be delineated in the subsequent two sections; following that, we will apply the technology to calculate the Green's functions in the Kerr black hole spacetime, up to first order in its mass and angular momentum. We will conclude with thoughts on possible future investigations.

\section{General Theory}

This section will summarize the key technical features of Green's functions we will need to understand in the development of perturbation theory in the following two sections. We refer the reader to Poisson's review \cite{PoissonReview:2003nc} for an in-depth discussion. We first examine the world function $\sigma_{x,x'}$, van Vleck determinant $\Delta_{x,x'}$ and the parallel propagator $g_{\mu\nu'}$, which are geometrical objects needed for the formal construction of the Green's functions themselves. We will record the equations obeyed by the Green's functions, and then describe the coefficients of $\delta[\sigma_{x,x'}]$ and $\Theta[\sigma_{x,x'}]$ in \eqref{HadamardForm_Scalar}, \eqref{HadamardForm_Photon} and \eqref{HadamardForm_Graviton}. Finally we will compute the $\sigma_{x,x'}$, $\Delta_{x,x'}$ and $g_{\mu\nu'}$ in Minkowski and perturbed Minkowski spacetimes.

{\bf World Function} \quad The world function $\sigma_{x,x'}$ defined in the introduction is half the square of the geodesic distance between $x$ and $x'$. Assuming there is a unique geodesic whose worldline has coordinates $\{\xi^\alpha[\lambda]|\lambda\in[0,1]; \xi^\alpha[0] = x'^\alpha, \ \xi^\alpha[1] = x^\alpha\}$, it has the integral representation
\begin{align}
\label{WorldFunction_Integral}
\sigma_{x,x'} = \frac{1}{2} \int_0^1 g_{\mu\nu}[\xi] \dot{\xi}^\mu \dot{\xi}^\nu \dd \lambda
\end{align}
with $\dot{\xi} \equiv \dd \xi/\dd \lambda$.

{\bf van Vleck Determinant} \quad Closely related to $\sigma_{x,x'}$ is the van Vleck determinant $\Delta_{x,x'}$
\begin{align}
\label{vanVleckDeterminant_Def}
\Delta_{x,x'} = -\frac{\det[ \partial_\mu \partial_{\nu'} \sigma_{x,x'} ]}{|gg'|^{1/2}}.
\end{align}

{\bf Parallel Propagator} \quad The parallel propagator $g_{\mu\nu'}$ is formed by contracting two sets of orthonormal basis tangent vector fields $\{\varepsilon^\mu_{\phantom{\mu}\text{A}}|\text{A},\mu = 0,1,2,3,\dots,d-1\}$, one based at $x$ and the other at $x'$. (The A-index is raised and lowered with $\eta_{\text{AB}}$ and the $\mu$-index is raised and lowered with the metric.)
\begin{align}
\label{ParallelPropagator_Def}
g_{\mu\nu'}[x,x'] \equiv \eta_{\text{AB}} \varepsilon_\mu^{\phantom{\mu}\text{A}}[x] \varepsilon_{\nu'}^{\phantom{\nu'}\text{B}}[x'],
\end{align}
with the boundary conditions that the metric be recovered at coincidence $x=x'$,
\begin{align}
\label{ParallelPropagator_BC}
g_{\mu\nu'}[x',x'] = g_{\mu'\nu'}[x'], \qquad
g_{\mu\nu'}[x,x] = g_{\mu\nu}[x].
\end{align}
The defining property of these vector fields $\{\varepsilon^\mu_{\phantom{\mu}\text{A}}\}$ and hence the parallel propagator itself, is that for a fixed pair of $x$ and $x'$, the $\{ \varepsilon^\mu_{\phantom{\mu}\text{A}} \}$ are parallel transported along the geodesic joining $x'$ to $x$. That is, $\dot{\xi}^\alpha \nabla_\alpha \varepsilon^\mu_{\phantom{\mu}\text{A}} = 0$ and consequently
\begin{align}
\label{ParallelPropagator_ParallelPropagated}
\dot{\xi}^\alpha \nabla_\alpha g_{\mu\nu'} = 0 .
\end{align}

{\bf Green's Function Equations} \quad Next we record the equations defining the Green's function. For the massless scalar,
\begin{align}
\label{GreensFunctionPDE_Scalar}
\Box_{x'} G_{x,x'} = \Box_x G_{x,x'} = \frac{\delta^d[x-x']}{|gg'|^{1/4}}
\end{align}
with $\Box_{x'} \equiv g^{\mu'\nu'} \nabla_{\mu'} \nabla_{\nu'}$ and $\Box_x \equiv g^{\mu\nu} \nabla_{\mu} \nabla_{\nu}$. For the Lorenz gauge photon ($\nabla^\alpha A_\alpha = 0$),\footnote{Our Christoffel symbol is $\Gamma^\alpha_{\mu\nu} = (1/2) g^{\alpha\lambda}(\partial_{\{\mu} g_{\nu\}\lambda} - \partial_\lambda g_{\mu\nu})$; Riemann tensor is $R^\alpha_{\phantom{a}\beta\mu\nu} = \partial_\mu \Gamma^\alpha_{\beta\nu} + \Gamma^\alpha_{\mu\sigma} \Gamma^\sigma_{\beta\nu} - \left( \mu \leftrightarrow \nu \right)$; the Ricci tensor and scalar $R_{\beta\nu} = R^\alpha_{\phantom{a}\beta\alpha\nu}$, $\mathcal{R} = g^{\beta\nu} R_{\beta\mu}$. Symmetrization is denoted, for example, by $T_{\{\alpha\beta\}} = T_{\alpha\beta} + T_{\beta\alpha}$. Antisymmetrization is denoted, for example, by $T_{[\alpha\beta]} = T_{\alpha\beta} - T_{\beta\alpha}$. Whenever we are performing an expansion in series of $h_{\mu\nu}$, the metric perturbation, indices of tensors are to be lowered and raised with the background metric $\gb_{\mu\nu}$.}
\begin{align}
\label{GreensFunctionPDE_Photon}
\Box_{x'} G_{\mu\nu'} - R_{\nu'}^{\phantom{\nu'}\lambda'} G_{\mu\lambda'}
    &= \Box_x G_{\mu\nu'} - R_\mu^{\phantom{\mu}\lambda} G_{\lambda\nu'} \nnn
    &= g_{\mu\nu'} \frac{\delta^d[x-x']}{|gg'|^{1/4}} .
\end{align}
DeWitt and Brehme \cite{DeWittBrehme:1960} points out that the divergence (with respect to $x$) of the Lorenz gauge photon Green's function is the negative gradient (with respect to $x'$) of the massless scalar Green's function
\begin{align}
\label{DivergenceOfPhotonG}
\nabla^\mu G_{\mu\nu'} = -\nabla_{\nu'} G_{x,x'} .
\end{align}
We will later note that our perturbative result satisfies \eqref{DivergenceOfPhotonG}. Proceeding to the de Donder gauge graviton ($\nabla^\mu \gamma_{\mu\nu} = \nabla_\nu \gamma/2$, with $\gamma \equiv g^{\mu\nu} \gamma_{\mu\nu}$),
{\allowdisplaybreaks\begin{align}
\label{GreensFunctionPDE_Graviton}
&\bigg(
\frac{1}{2} \left( \delta^\mu_{\{\alpha} \delta^\nu_{\beta\}} - g_{\alpha\beta} g^{\mu\nu} \right) \left( \Box - \mathcal{R} + 2\Lambda \right)
    + 2 R^{\mu\phantom{\alpha}\nu\phantom{\beta}}_{\phantom{\mu}\alpha\phantom{\nu}\beta} \nnn
&\qquad \qquad + R^\nu_{\phantom{\nu}\{\alpha} \delta^\mu_{\beta\}} - g^{\mu\nu} R_{\alpha\beta} - g_{\alpha\beta} R^{\mu\nu} \bigg) G_{\mu\nu \rho'\sigma'} \nnn
&\qquad \qquad = \delta_{\alpha\beta;\rho'\sigma'} \frac{\delta^d[x-x']}{|gg'|^{1/4}}
\end{align}}
where we have included a non-zero cosmological constant $\Lambda$. The $\delta_{\alpha\beta;\sigma'\rho'}$ is built out of the parallel propagator $g_{\mu\nu'}[x,x']$,
\begin{align}
\delta_{\alpha\beta;\rho'\sigma'} \equiv \frac{1}{2} \left( g_{\alpha\rho'} g_{\beta\sigma'} + g_{\alpha\sigma'} g_{\beta\rho'} \right) .
\end{align}
Green's functions are bitensors. Coordinate transformations at $x$ can be carried out independently from $x'$ (and vice versa). Derivatives with respect to $x$ are independent of that with respect to $x'$, so for instance, $\nabla_\mu G_{\alpha\beta'} = \partial_\mu G_{\alpha\beta'} - \Gamma^\lambda_{\mu\alpha} G_{\lambda\beta'}$.

{\bf Hadamard form} \quad We now have sufficient vocabulary to describe the coefficients of $\delta[\sigma_{x,x'}]$ and $\Theta[\sigma_{x,x'}]$ in the Green's functions in equations \eqref{HadamardForm_Scalar}, \eqref{HadamardForm_Photon} and \eqref{HadamardForm_Graviton}. Assuming $x$ and $x'$ lie in a region of spacetime where there is a unique geodesic joining them, in 4 dimensional spacetimes, the null cone pieces are built out of the van Vleck determinant and the parallel propagators
{\allowdisplaybreaks\begin{align}
\label{NullConeFunction_Scalar}
U_{x,x'}                    &= \sqrt{\Delta_{x,x'}} \\
\label{NullConeFunction_Photon}
U_{\mu\nu'}                 &= \sqrt{\Delta_{x,x'}} g_{\mu\nu'} \\
\label{NullConeFunction_Graviton}
U_{\mu\nu \alpha'\beta'}    &= \sqrt{\Delta_{x,x'}} P_{\mu\nu \alpha'\beta'}
\end{align}}
where
\begin{align}
P_{\mu\nu \alpha'\beta'} \equiv \frac{1}{2} \left( g_{\mu\alpha'} g_{\nu\beta'} + g_{\mu\beta'} g_{\nu\alpha'} - g_{\mu\nu} g_{\alpha'\beta'} \right) .
\end{align}
The tail portions of the Green's functions satisfy the homogeneous equations, for example, $\Box_x V_{x,x'} = \Box_{x'} V_{x,x'} = 0$; Poisson \cite{PoissonReview:2003nc} explains the appropriate non-trivial boundary conditions the tail function $V$s must satisfy. Moreover, the derivation of \eqref{NullConeFunction_Scalar}, \eqref{NullConeFunction_Photon} and \eqref{NullConeFunction_Graviton} shows that the geometric tensors in the wave equation for photons and gravitons only contribute to the tail portion of the field propagation; while it is the differential operator, namely $\Box$, that contributes to both the behavior of the null propagation and that of the tail piece. We will also witness this in the perturbative framework we are about to pursue.

It is appropriate at this point to highlight that these geometrical constructs, from which the light cone piece of the Green's functions are built, have physical meaning for the cosmologist. For example, the world function obeys the following equation involving the van Vleck determinant
\begin{align}
\label{ExpansionEquation}
\Box_x \sigma_{x,x'} + \nabla^\mu \sigma_{x,x'} \nabla_\mu \ln \Delta_{x,x'} = d .
\end{align}
Because $\nabla^\alpha \sigma_{x,x'}$ is proportional to the tangent vector $\dot{\xi}^\alpha$ at $x$ (it points in the direction of greatest rate of change in geodesic distance), $\Box_x \sigma_{x,x'} \propto \nabla_\alpha \dot{\xi}^\alpha$ describes the rate of change of the cross sectional area of the congruence of geodesics (the ``expansion") through the neighborhood of $x$, which via \eqref{ExpansionEquation} is related to the gradient of $\Delta_{x,x'}$ along the geodesics. This expansion scalar is related to the evolution of the angular diameter distance, which then in turn is related to the luminosity distance relation. (See, for example. Visser \cite{Visser:1992pz} and Flanagan et al. \cite{Flanagan:2008kz}.) Along similar lines, initially parallel null rays from an extended source become deflected due to gravitational effects (weak lensing). Since the parallel propagator describes the parallel transport of an orthonormal reference frame along these trajectories, namely
\begin{align}
g^\mu_{\phantom{\mu}\nu'}[x,x'] \varepsilon^{\nu'}_{\phantom{\nu'}\text{A}}[x'] = \varepsilon^\mu_{\phantom{\mu}\text{A}}[x] \quad \text{(see \eqref{ParallelPropagator_Def})},
\end{align}
they ought to contain physical content regarding polarization, rotation and shear of null bundles of photons. To sum, the light cone part of the massless scalar and photon Green's function should provide an alternate means, from the standard ones in use by cosmologists today, of getting at the physics of null light traveling through the universe. This warrants more study.

Before moving on to develop our perturbation theory, let us take a few moments to calculate the world function, van Vleck determinant and parallel propagator up to first order in $h_{\mu\nu}$ in perturbed Minkowski spacetime. This will allow us to construct the null cone piece of the scalar, photon and graviton Green's function, and in turn, serve as a consistency check on our first Born approximation results below.\footnote{Some of the results here can be found in Kovacs and Thorne \cite{KovacsThorne:1975} and Pfenning and Poisson \cite{PfenningPoisson:2000zf}, but we include them so that the discussion is self-contained.} In fact, this was how Kovacs and Thorne \cite{KovacsThorne:1975} constructed the null cone piece of their Green's functions, by calculating separately the van Vleck determinant and Synge's world function. But we shall argue that this is not necessary. The Born series scheme we have devised gives us a single coherent framework where all three geometric objects appearing in the null cone piece of the Green's function are byproducts of the computation. Specifically, the van Vleck determinant and the world function can be read off the massless scalar Green's function $G_{x,x'}$, and the parallel propagator can be read off the Lorenz gauge photon Green's function $G_{\mu\nu'}$.

{\bf $\sigma$, $\Delta$ and $g_{\mu\nu'}$ in Minkowski} \quad The geodesic equation in Minkowski spacetime is
\begin{align}
\frac{\dd^2 \bar{\xi}^\alpha}{\dd \lambda^2} = 0,
\end{align}
with boundary conditions $\bar{\xi}^\alpha[0] = x'^\alpha$ and $\bar{\xi}^\alpha[1] = x^\alpha$. The solution is
\begin{align}
\bar{\xi}^\alpha[\lambda] = x'^\alpha + \lambda (x-x')^\alpha .
\end{align}
Inserting this into \eqref{WorldFunction_Integral} yields the world function
\begin{align}
\bar{\sigma}_{x,x'} = \frac{1}{2} \eta_{\mu\nu} \Delta^\mu \Delta^\nu ,
\end{align}
where we have defined
\begin{align}
\Delta^\mu \equiv (t-t', \vec{x}-\vec{x}')^\mu,
\end{align}
which is not to be confused with the van Vleck determinant (we will always place the spacetime coordinates as subscripts for the latter). Since
{\allowdisplaybreaks\begin{align}
\label{Derivatives}
\partial_\mu \bar{\sigma}_{x,x'} = \Delta_\mu, \quad
\partial_{\mu'} \bar{\sigma}_{x,x'} = -\Delta_\mu \nnn
\partial_\mu \partial_{\nu'} \bar{\sigma}_{x,x'} = -\eta_{\mu\nu}
\end{align}}
by equation \eqref{vanVleckDeterminant_Def}, the van Vleck determinant is unity. In Cartesian coordinates, the parallel propagator is numerically equal, component-by-component, to the Minkowski metric $\eta_{\mu\nu}$.
\begin{align}
\overline{\Delta}_{x,x'} = 1, \qquad \bar{g}_{\mu\nu'} = \eta_{\mu\nu} .
\end{align}
We shall soon be making heavy use of the massless scalar $\bar{G}_{x,x'}$, photon $\bar{G}_{\mu\nu'}$, and graviton $\bar{G}_{\mu\nu \alpha'\beta'}$ Green's functions in 4 dimensional Minkowski spacetime, so let us record their explicit expressions here
{\allowdisplaybreaks\begin{align}
\label{GreensFunctionMinkowski_Scalar}
\bar{G}_{x,x'} &= \frac{\Theta[t-t']\delta[\bar{\sigma}_{x,x'}]}{4\pi} \nnn
&= \frac{\Theta[t-t'] \delta[t-t'-|\vec{x}-\vec{x}'|]}{4\pi|\vec{x}-\vec{x}'|} \\
\label{GreensFunctionMinkowski_Photon}
\bar{G}_{\alpha\beta'} &= \eta_{\alpha\beta} \bar{G}_{x,x'} \\
\label{GreensFunctionMinkowski_Graviton}
\bar{G}_{\mu\nu \alpha'\beta'} &= \bar{P}_{\mu\nu \alpha\beta} \bar{G}_{x,x'}
\end{align}}
with
\begin{align}
\bar{P}_{\mu\nu \alpha\beta} = \frac{1}{2}\left( \eta_{\mu\{\alpha} \eta_{\beta\}\nu} - \eta_{\mu\nu}\eta_{\alpha\beta} \right) .
\end{align}
The photon here obeys the Lorenz gauge $\eta^{\mu\nu} \partial_\mu A_\nu = 0$ while the graviton the de Donder gauge $\partial^\mu h_{\mu\nu} = \partial_\nu \eta^{\alpha\beta}h_{\alpha\beta}/2$. For computational purposes, we record that the $\bar{P}$ has the following symmetries
\begin{align}
\label{PbarSymmetries}
\bar{P}_{\mu\nu \alpha\beta} = \bar{P}_{\alpha\beta \mu\nu} = \bar{P}_{\nu\mu \alpha\beta} = \bar{P}_{\mu\nu \beta\alpha} .
\end{align}

{\bf $\sigma$, $\Delta$ and $g_{\mu\nu'}$ in perturbed Minkowski} \quad To tackle these geometric entities in perturbed Minkowski, we start by noting that the integral in \eqref{WorldFunction_Integral} defines a variational principle for geodesics. For fixed end points $x'$ and $x$, and $\lambda$ an affine parameter, the paths which extremizes the integral in \eqref{WorldFunction_Integral} are the geodesics. Let $\xi$ be the geodesic in perturbed Minkowski spacetime joining $x'$ to $x$. If we were to solve it perturbatively, we can try $\xi = \bar{\xi} + \delta \xi$, where $\delta \xi$ can be viewed as a small displacement, and plug this ansatz into the integral in \eqref{WorldFunction_Integral}. But since the integral defines a variational principle, that means the first order variation of the integrand, due to the $\mathcal{O}[\delta \xi]$ deviation of the geodesic from the Minkowski one, is zero. To first order in $h_{\mu\nu}$, the world function can thus be obtained from \eqref{WorldFunction_Integral} by simply setting $\xi = \bar{\xi}$.
\begin{align}
\label{WorldFunction_PerturbedMinkowski}
\sigma_{x,x'} \approx \bar{\sigma}_{x,x'} + \Delta^\mu \Delta^\nu \widehat{\mathcal{I}}^{(0)}_{\mu\nu} ,
\end{align}
where
\begin{align}
\label{I0}
\widehat{\mathcal{I}}^{(0)}_{\mu\nu} \equiv \frac{1}{2} \int_0^1 h_{\mu\nu}[\bar{\xi}] \dd \lambda .
\end{align}
(The reason for the name $\widehat{\mathcal{I}}^{(0)}_{\mu\nu}$ will be clear later.) Now put \eqref{WorldFunction_PerturbedMinkowski} into \eqref{vanVleckDeterminant_Def}, and employ \eqref{Derivatives}. Then use the following relation, that for matrices $A$ and $B$ such that $B$ is a small perturbation relative to $A$,
\begin{align}
\det[A + B] = \det [A] \left( 1 + \text{Tr}[A^{-1} B] + \dots \right) .
\end{align}
(Tr denotes trace, and $A^{-1}$ is the inverse of $A$.) We then deduce the square root of the van Vleck determinant is
\begin{align}
\label{vanVleckDeterminant_PerturbedMinkowski}
&\sqrt{\Delta_{x,x'}} \approx
\bigg( 1 - \frac{1}{4} h - \frac{1}{4} h'
    + \widehat{\mathcal{I}}^{(0)} \\
&\qquad \qquad
    + \Delta^\beta \left( \partial^\mu - \partial^{\mu'} \right) \widehat{\mathcal{I}}^{(0)}_{\mu\beta}
    -\frac{1}{2} \Delta^\alpha \Delta^\beta \partial^\mu \partial_{\mu'} \widehat{\mathcal{I}}^{(0)}_{\alpha\beta} \bigg) . \nn
\end{align}
Here, $h' \equiv \eta^{\mu\nu} h_{\mu'\nu'}$ and $\widehat{\mathcal{I}}^{(0)} \equiv \eta^{\mu\nu} \widehat{\mathcal{I}}^{(0)}_{\mu\nu}$.

In \cite{Visser:1992pz}, Visser developed perturbation theory for solving the van Vleck determinant. In particular, he showed that the $\mathcal{O}[h]$ accurate $\sqrt{\Delta_{x,x'}}$ is given by (his equation 61)
\begin{align}
\label{vanVleckDeterminant_PerturbedMinkowski_RicciForm}
\sqrt{\Delta_{x,x'}} \approx 1 + \frac{\Delta^\alpha \Delta^\beta}{2} \int_0^1 \dd \lambda (1-\lambda)\lambda (R|1)_{\alpha\beta}[\bar{\xi}],
\end{align}
where $(R|1)_{\alpha\beta}[\bar{\xi}]$ is the linearized Ricci tensor evaluated on the unperturbed geodesic $\bar{\xi}$.

Let us show the equivalence of \eqref{vanVleckDeterminant_PerturbedMinkowski} and \eqref{vanVleckDeterminant_PerturbedMinkowski_RicciForm}. First, we write down the explicit form of the linearized Ricci tensor in Cartesian coordinates. One is lead to the expression
\begin{align}
\sqrt{\Delta_{x,x'}} &\approx 1 + \frac{\Delta^\alpha \Delta^\beta}{2} \int_0^1 \dd \lambda
    (1-\lambda)\lambda \bigg( \partial^{\mu''} \partial_{\alpha''} h_{\beta''\mu''} \nonumber\\
    &- \frac{1}{2} \partial_{\alpha''} \partial_{\beta''} h \bigg)
    - \frac{1}{2} \Delta^\alpha\Delta^\beta \partial^\mu \partial_{\mu'} \widehat{\mathcal{I}}^{(0)}_{\alpha\beta},
\end{align}
where we have employed $(1-\lambda)\partial_{\mu''} = \partial_{\mu'}$ and $\lambda \partial_{\mu''} = \partial_{\mu}$. In the first line, the $\Delta^\alpha \partial_{\alpha''} = \dd/\dd\lambda$. This can be integrated-by-parts, and the resulting $(1-2\lambda) \partial^{\mu''}$ is $\partial^{\mu'}-\partial^\mu$, and can be pulled out of the integral,
\begin{align}
\frac{\Delta^\alpha \Delta^\beta}{2} \int_0^1 \dd \lambda
    (1-\lambda)\lambda \partial^{\mu''} \partial_{\alpha''} h_{\beta''\mu''}
= \Delta^\beta (\partial^\mu-\partial^{\mu'}) \widehat{\mathcal{I}}^{(0)}_{\mu\beta}.
\end{align}
What remains is to demonstrate that
\begin{align*}
-\frac{\Delta^\alpha \Delta^\beta}{4} \int_0^1 \dd \lambda
    (1-\lambda)\lambda \partial_{\alpha''} \partial_{\beta''} h = -\frac{1}{4} h -\frac{1}{4} h' + \widehat{\mathcal{I}}^{(0)}.
\end{align*}
This relation can be reached by recognizing $\Delta^\alpha \Delta^\beta \partial_{\alpha''} \partial_{\beta''} h = \dd^2 h/\lambda^2$, followed by integrating-by-parts the $\dd^2/\dd\lambda^2$.

Equation \eqref{ParallelPropagator_ParallelPropagated} says the parallel propagator is parallel propagated along $\xi$. If we write $g_{\mu\nu'} = \eta_{\mu\nu} + h_{\mu\nu'}$ and keep only the $\mathcal{O}[h]$ terms in the Christoffel symbol in $\nabla_\alpha h_{\mu\nu'} = \partial_\alpha h_{\mu\nu'} - \Gamma^\lambda_{\alpha\mu} h_{\lambda\nu'}$, \eqref{ParallelPropagator_ParallelPropagated} is then approximately equivalent to
\begin{align}
\label{ParallelPropagator_ODE_PerturbedMinkowski}
&\frac{\dd}{\dd \lambda} h_{\mu\nu'}[\xi[\lambda],x'] \\
& \ = \frac{1}{2} \left( \partial_{\rho''} h_{\mu''\nu''}[\bar{\xi}] + \partial_{\mu''} h_{\rho''\nu''}[\bar{\xi}] - \partial_{\nu''} h_{\mu''\rho''}[\bar{\xi}] \right) \dot{\xi}^\rho[\lambda] \nn
\end{align}
where the derivatives are with respect to $\bar{\xi}$; for example, $\partial_{\mu''} \equiv \partial/\partial \bar{\xi}^\mu$. Since $\delta \xi$ has to begin at $\mathcal{O}[h]$, that means to the first order, we can replace $\dot{\xi}^\rho$ with $\dot{\bar{\xi}}^\rho = \Delta^\rho$. Recognizing
\begin{align}
\label{DerivativeLambda}
\frac{\dd h_{\mu''\nu''}[\bar{\xi}]}{\dd \lambda} = \Delta^\rho \partial_{\rho''} h_{\mu''\nu''}[\bar{\xi}]
\end{align}
and recalling the boundary conditions \eqref{ParallelPropagator_BC} then allow us to integrate \eqref{ParallelPropagator_ODE_PerturbedMinkowski} to deduce
\begin{align}
\label{ParallelPropagator_PerturbedMinkowski}
g_{\mu\nu'} \approx \eta_{\mu\nu} + \frac{1}{2} \left( h_{\mu\nu} + h_{\mu'\nu'} \right)
    + \frac{\Delta^\rho}{2} \int_0^1 \partial_{[\mu''} h_{\nu'']\rho''}[\bar{\xi}] \dd \lambda .
\end{align}
As can be checked explicitly,
\begin{align}
\partial_{\mu''} h_{\nu''\rho''}[\bar{\xi}] = (\partial_\mu + \partial_{\mu'}) h_{\nu''\rho''}.
\end{align}
We may thus re-write \eqref{ParallelPropagator_PerturbedMinkowski} in terms of $\widehat{\mathcal{I}}^{(0)}_{\mu\nu}$ in \eqref{I0},
\begin{align}
\label{ParallelPropagator_PerturbedMinkowski_I0Form}
g_{\mu\nu'} \approx \eta_{\mu\nu} + \frac{1}{2} \left( h_{\mu\nu} + h_{\mu'\nu'} \right)
    + \Delta^\rho ( \partial_{[\mu} + \partial_{[\mu'} ) \widehat{\mathcal{I}}^{(0)}_{\nu]\rho} .
\end{align}
To be clear, $h_{\mu\nu}$ and $h_{\mu'\nu'}$ are the metric perturbations at $x$ and $x'$ respectively; while $\eta_{\mu\nu} + h_{\mu\nu'}$ is the parallel propagator in perturbed Minkowski spacetime.

\section{Perturbation Theory}

We now describe the Born series method to solve the Green's functions in a formal power series in $h_{\mu\nu}$, the metric perturbation.

\begin{figure}
\includegraphics[width=3.5in]{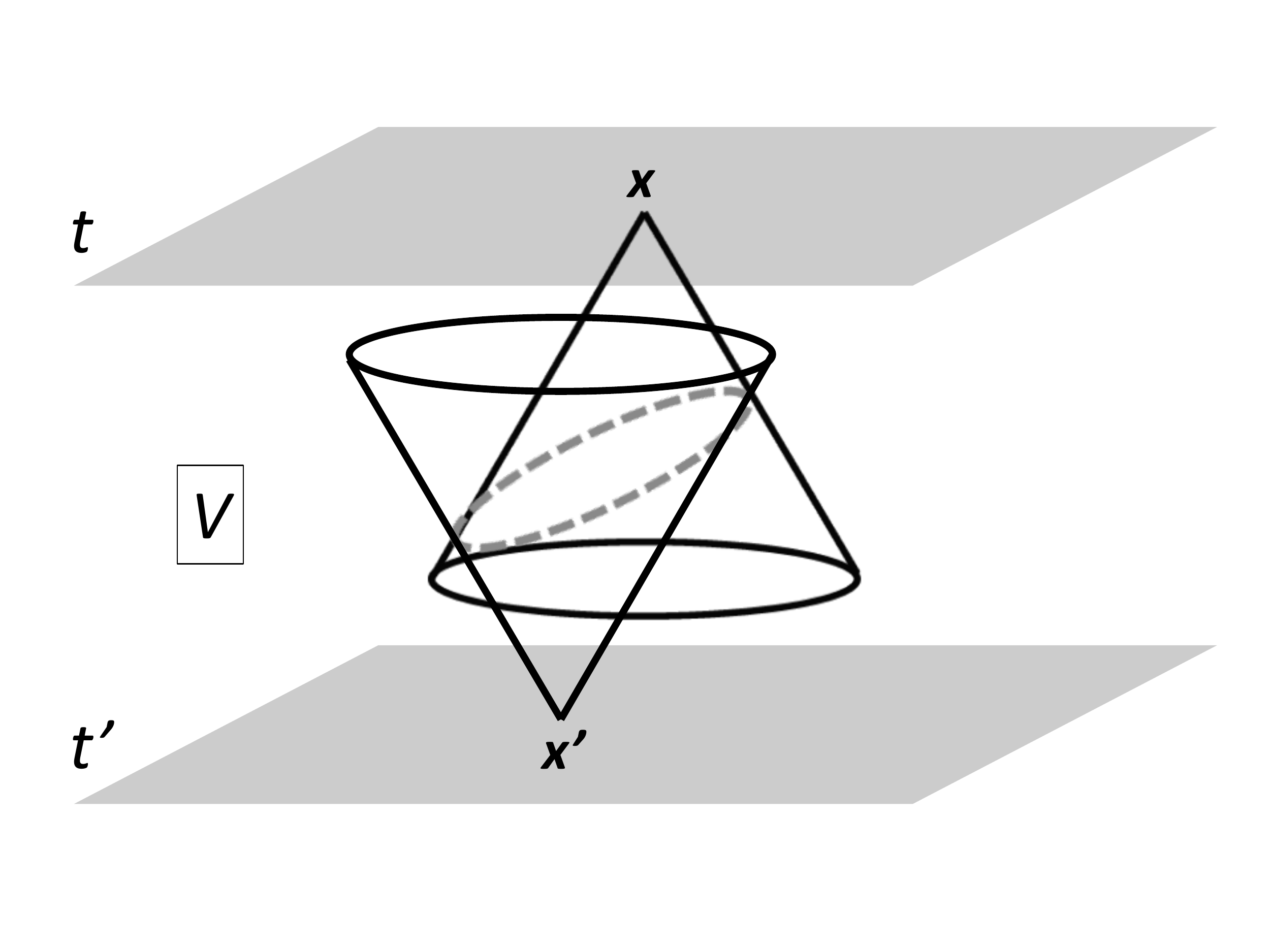} \\
\includegraphics[width=3.5in]{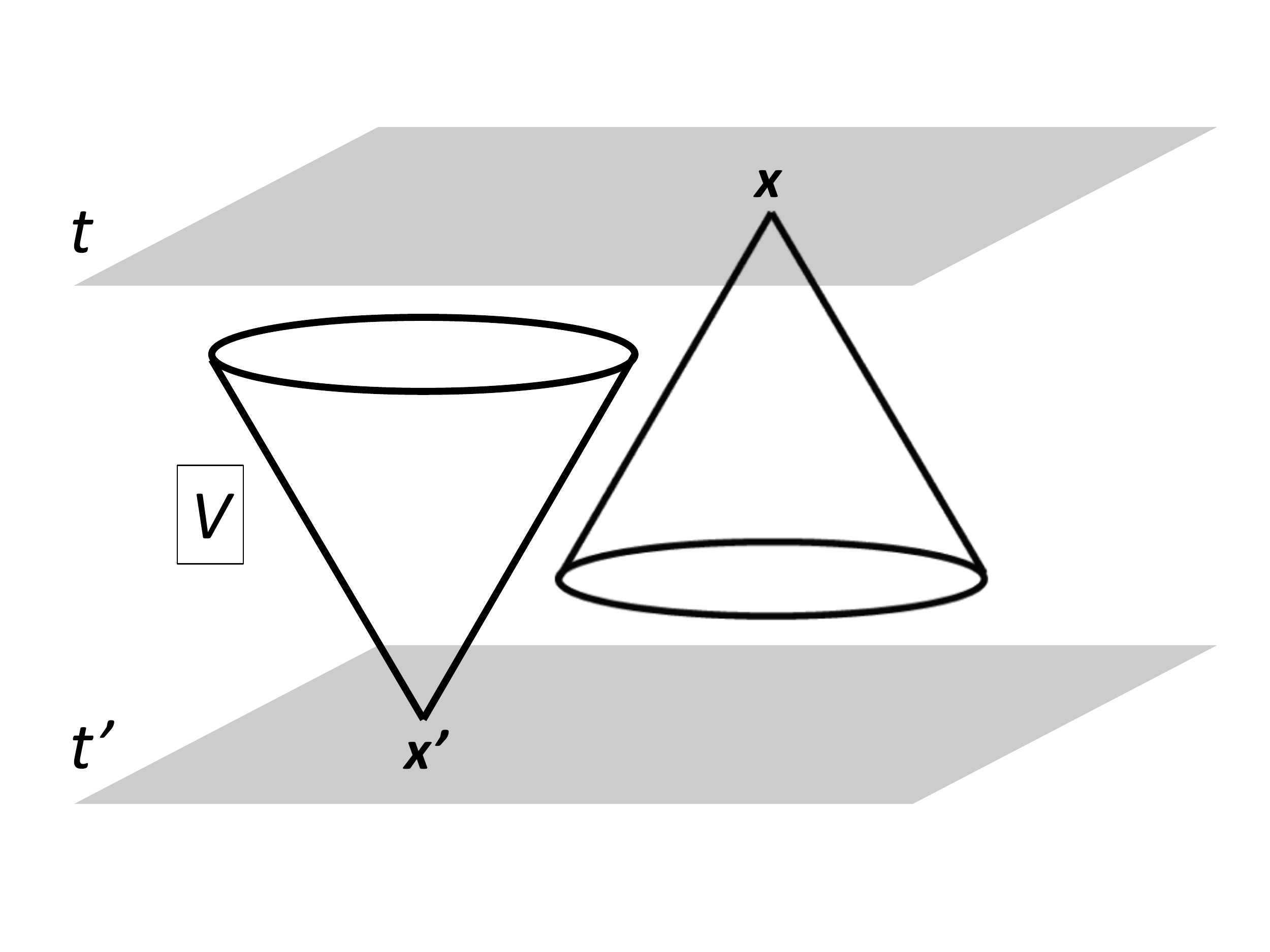}
\caption{\emph{Top Panel}: The intersection of the interiors of the future null cone of $x'$ and that of the past null cone of $x$ always defines a finite (as opposed to infinite) region of spacetime. Moreover, if (and only if) $x'$ lies on or within the interior of the backward light cone of $x$ (or, equivalently, if and only if $x$ lies on or within the interior of the forward light cone of $x'$), then there is a non-trivial intersection (indicated by the dark dashed oval) between the forward light cone of $x'$ and backward light cone of $x$, which in 3-space we shall show is a prolate ellipsoid, when the background is Minkowski.
\emph{Bottom Panel}: If $x'$ lies outside the backward light cone of $x$ (or, equivalently, if $x$ lies outside the forward light cone of $x'$), then there is no intersection between the forward light cone of $x'$ and backward light cone of $x$.}
\label{TwoLightConesIntersectFigure}
\end{figure}

{\bf Scalar} \quad The quadratic action of the minimally coupled massless scalar field evaluated in the perturbed metric $g_{\mu\nu} = \gb_{\mu\nu} + h_{\mu\nu}$ reads
\begin{align}
S_\varphi[g] \equiv \frac{1}{2} \int \dd^d x'' |g''|^{\frac{1}{2}} \nabla^{\alpha''} \varphi \nabla_{\alpha''} \varphi
\end{align}
while the same action evaluated in the background metric $\gb_{\mu\nu}$, with $\Db_\mu$ denoting the covariant derivative with respect to it, is
\begin{align}
S_\varphi[\gb] \equiv \frac{1}{2} \int \dd^d x'' |\gb''|^{\frac{1}{2}} \Db^{\alpha''} \varphi \Db_{\alpha''} \varphi .
\end{align}
In $S_\varphi[g]$, if we replace one field with $\bG_{x,x''}$, the Green's function in $\gb_{\mu\nu}$, and the other with $G_{x'',x'}$, the Green's function in $g_{\mu\nu}$, upon integration-by-parts, and using \eqref{GreensFunctionPDE_Scalar}, we see that
\begin{align}
\label{TwiceAction_Scalar_g}
&2 S_\varphi[g;\bG_{x,x''}, G_{x'',x'}] \\
&= \int \dd^d x'' |g''|^{\frac{1}{2}} \nabla^{\alpha''} \bG_{x,x''} \nabla_{\alpha''} G_{x'',x'}
    = - \bG_{x,x'} . \nn
\end{align}
Similarly, by replacing one of the fields in $S_\varphi[\gb]$ with $\bG_{x'',x'}$ and the other with $G_{x,x''}$, one obtains
\begin{align}
\label{TwiceAction_Scalar_gbar}
&2 S_\varphi[\gb;G_{x,x''}, \bG_{x'',x'}] \\
&= \int \dd^d x'' |\gb''|^{\frac{1}{2}} \Db^{\alpha''} G_{x,x''} \Db_{\alpha''} \bG_{x'',x'}
    = - G_{x,x'} . \nn
\end{align}
The surface terms incurred during integration-by-parts in \eqref{TwiceAction_Scalar_g} and \eqref{TwiceAction_Scalar_gbar} are zero because the surface integrands at hand, namely $\bG_{x,x''} \nabla^{\alpha''} G_{x'',x'}$ and $G_{x,x''} \Db^{\alpha''} \bG_{x'',x'}$, due the causal structure of the Green's functions, are non-zero only in the spacetime region defined by the intersection of the interiors of the past light cone of $x$ with that of the future null cone of $x'$. As Fig. \eqref{TwoLightConesIntersectFigure} informs us, this intersection is always a finite region of spacetime. As long as we are dealing with a spacetime manifold that is infinite (or semi-infinite) in extent, this finite region of intersection lies deep inside the region enclosed by the surface at infinity, and hence does not contribute to the surface integral itself.
Subtracting the equations \eqref{TwiceAction_Scalar_g} and \eqref{TwiceAction_Scalar_gbar} then hands us an integral equation for $G_{x,x'}$:
{\allowdisplaybreaks\begin{align}
\label{GreensFunctionIntegralEquation_Scalar}
&G_{x,x'} - \bG_{x,x'} \\
&= \int \dd^d x'' |g''|^{\frac{1}{2}} g^{\alpha''\beta''} \nabla_{\alpha''} \bG_{x,x''} \nabla_{\beta''} G_{x'',x'} \nnn
&\qquad - \int \dd^d x'' |\gb''|^{\frac{1}{2}} \gb^{\alpha''\beta''} \Db^{\alpha''} G_{x,x''} \Db_{\alpha''} \bG_{x'',x'} . \nn
\end{align}}
\emph{First Born Approximation} \quad Perturbation theory may now be carried out by iterating \eqref{GreensFunctionIntegralEquation_Scalar} as many times as one wishes (followed by dropping the remainder integral terms containing $G_{x,x'}$), and expanding
\begin{align}
\label{MetricExpansion_gb+h_I}
g^{\alpha\beta} &= \gb^{\alpha\beta} - h^{\alpha\beta} + \dots \\
\label{MetricExpansion_gb+h_II}
|g|^{\frac{1}{2}} &= |\gb|^{\frac{1}{2}} \left( 1 + \frac{1}{2} h + \dots \right); \ h \equiv \gb^{\alpha\beta} h_{\alpha\beta}
\end{align}
to as high an order in $h_{\mu\nu}$ as desired. (We are now raising and lowering all indices with the background metric $\gb_{\mu\nu}$.) To obtain the first Born approximation, the $\mathcal{O}[h]$-accurate result for $G_{x,x'}$, one replaces the $G_{x,x''}$ occurring within the integrals in \eqref{GreensFunctionIntegralEquation_Scalar} with $\bar{G}_{x,x''}$ and only need to expand the $|g''|^{1/2}$ and $g^{\alpha''\beta''}$ to first order. The result is
\begin{align}
\label{GreensFunctionBornApproximation_Scalar_gb}
&G_{x,x'} \approx \bar{G}_{x,x'}
+ \int\dd^d x'' |\gb''|^{\frac{1}{2}} \\
&\qquad \times
    \left\{ \partial_{\alpha''} \bar{G}_{x,x''} \left(\frac{1}{2} h'' \bar{g}^{\alpha''\beta''} - h^{\alpha''\beta''} \right) \partial_{\beta''} \bar{G}_{x'',x'} \right\} \nn
\end{align}
with $h'' \equiv \gb^{\rho''\tau''} h_{\rho''\tau''}$. In perturbed Minkowski spacetime, we set $\gb_{\mu\nu} = \eta_{\mu\nu}$, employ Cartesian coordinates, and then use the spacetime translation symmetry reflected by the Green's function $\bG_{x,x'}$ for any $d$, namely
\begin{align}
\label{Minkowski_TranslationSymmetry}
\partial_\mu \bG_{x,x'} = -\partial_{\mu'} \bG_{x,x'} ,
\end{align}
to pull the two derivatives out of the integral
\begin{align}
\label{GreensFunctionBornApproximation_Scalar_eta}
&G_{x,x'} \approx \bar{G}_{x,x'} \\
&+ \partial_{\alpha} \partial_{\beta'}
    \int\dd^d x'' \bar{G}_{x,x''} \left(\frac{1}{2} h'' \eta^{\alpha\beta} - h^{\alpha''\beta''} \right) \bar{G}_{x'',x'} \nn
\end{align}
with $h'' \equiv \eta^{\rho\tau} h_{\rho''\tau''}$. This matches equation 2.27 of DeWitt and DeWitt \cite{DeWittDeWitt:1964}, if we note that their Green's function is negative of ours.

{\bf Photon} \quad Next, we turn to the photon. The Maxwell action in terms of electric and magnetic fields $F_{\mu\nu}$ is
\begin{align}
S_\text{Maxwell} = -\frac{1}{4} \int \dd^d x'' |g''|^{\frac{1}{2}} g^{\mu''\alpha''} g^{\nu''\beta''} F_{\mu''\nu''} F_{\alpha''\beta''} .
\end{align}
We have already noted in the introduction, that this action $S_\text{Maxwell}$ enjoys a conformal symmetry in 4 dimensions, namely, it evaluates to the same object in both the metric $g_{\mu\nu}$ and the metric $a^2 g_{\mu\nu}$; the conformal factor $a^2$ drops out. Whenever there is such a conformal factor, for instance, as in the context of a spatially flat inhomogeneous FLRW universe described by the metric in \eqref{FLRWMetric_Perturbed} we will choose the Lorenz gauge with respect to $g_{\mu\nu}$ and not $a^2 g_{\mu\nu}$:
\begin{align}
\label{Photon_LorenzGauge}
\nabla^\mu A_\mu \equiv \frac{1}{|g|^{\frac{1}{2}}} \partial_\mu \left( |g|^{\frac{1}{2}} g^{\mu\nu} A_\nu \right) = 0
\end{align}
so that the dynamics of $A_\mu$ will also be blind to $a^2$. The quadratic action for the photon's vector potential $A_\mu$ evaluated in the metric $g_{\mu\nu} = \gb_{\mu\nu} + h_{\mu\nu}$ is
\begin{align}
S_A[g]
    &= -\frac{1}{2} \int \dd^d x'' |g''|^{\frac{1}{2}} \bigg(
            \nabla^{\alpha''} A^{\beta''} \nabla_{\alpha''} A_{\beta''} \\
            &\qquad \qquad \qquad \qquad
                + R^{\alpha''\beta''} A_{\alpha''} A_{\beta''} \bigg) . \nn
\end{align}
Via steps analogous to the ones taken to obtain the integral equation for the scalar Green's function, replacing one field with $G$ and the other with $\bG$, we can write down the corresponding integral equation for the photon Green's function $G_{\mu\nu'}$ in the perturbed spacetime $g_{\mu\nu} = \gb_{\mu\nu} + h_{\mu\nu}$:
{\allowdisplaybreaks\begin{align}
\label{GreensFunctionIntegralEquation_Photon}
&G_{\mu\nu'} - \bG_{\mu\nu'} \\
&= \int \dd^d x'' |g''|^{\frac{1}{2}} \bigg(
    g^{\alpha''\beta''} g^{\lambda''\rho''} \nabla_{\alpha''} \bG_{\mu\lambda''} \nabla_{\beta''} G_{\rho''\nu'} \nnn
    &\qquad \qquad \qquad \qquad
    + R^{\lambda''\rho''} \bG_{\mu\lambda''} G_{\rho''\nu'} \bigg) \nnn
&- \int \dd^d x'' |\gb''|^{\frac{1}{2}} \bigg(
    \gb^{\alpha''\beta''} \gb^{\lambda''\rho''} \Db_{\alpha''} G_{\mu\lambda''} \Db_{\beta''} \bG_{\rho''\nu'} \nnn
    &\qquad \qquad \qquad \qquad
        + \bar{R}^{\lambda''\rho''} G_{\mu\lambda''} \bG_{\rho''\nu'} \bigg) . \nn
\end{align}}
Here and below, the barred geometric tensors such as $\bar{R}_{\mu\nu}$ are built out of $\gb_{\mu\nu}$; whereas the un-barred ones are built out of $g_{\mu\nu}$.

\emph{First Born Approximation} \quad Like the scalar case, one may now pursue perturbation theory of the photon Green's function by iterating the integral equation \eqref{GreensFunctionIntegralEquation_Photon} however many times (followed by dropping the remainder integral terms containing $G_{\mu\nu'}$) and perform the expansion in \eqref{MetricExpansion_gb+h_I} and \eqref{MetricExpansion_gb+h_II}, and of the Christoffel symbols
\begin{align}
\label{ChristoffelExpansion_gb+h}
&\Gamma^\mu_{\alpha\beta}[g] - \Gamma^\mu_{\alpha\beta}[\gb] \\
&= \frac{1}{2} (\gb^{\mu\lambda}-h^{\mu\lambda}+\dots)\left( \Db_{\{\alpha} h_{\beta\}\lambda}-\Db_\lambda h_{\alpha\beta} \right) \nn
\end{align}
to whatever order in $h_{\mu\nu}$ one wishes. To $\mathcal{O}[h]$, we merely need to replace the $G_{\mu\nu'}$ occurring under the integral sign in \eqref{GreensFunctionIntegralEquation_Photon} with $\bG_{\mu\nu'}$ and develop the necessary expansion to linear order in $h_{\mu\nu}$. The additional complication in the photon case here, and the graviton case below, is that one has to deal with integrals of the schematic form $\int \bG (\Gamma|1) \Db \bG$, arising from the covariant differentiation of the Green's functions. The $(\Gamma|1)$ is the first order in $h_{\mu\nu}$ variation of the Christoffel symbol,
\begin{align}
(\Gamma|1)^\mu_{\alpha\beta}
&= \frac{1}{2} \gb^{\mu\lambda} \left( \Db_{\{\alpha} h_{\beta\}\lambda}-\Db_\lambda h_{\alpha\beta} \right) .
\end{align}
For such terms, we will choose to integrate by parts, moving all the (single) derivatives acting on the $h_{\mu\nu}$s in the $(\Gamma|1)$ onto the un-perturbed Green's functions $\bG$. (As already argued, there are no surface terms.) The ensuing manipulations require the use of equations \eqref{GreensFunctionPDE_Photon} and \eqref{DivergenceOfPhotonG}. About a generic perturbed spacetime $g_{\mu\nu} = \gb_{\mu\nu} + h_{\mu\nu}$, we then gather that
\begin{widetext}
{\allowdisplaybreaks\begin{align}
\label{GreensFunctionBornApproximation_Photon_gb+h}
G_{\mu\nu'}
&\approx \bar{G}_{\mu\nu'}
+ \frac{1}{2} \bar{G}_{\mu\alpha'} h^{\alpha'}_{\phantom{\alpha'}\nu'}
+ \frac{1}{2} h_{\mu}^{\phantom{\mu}\alpha}
\bar{G}_{\alpha\nu'} \nnn
&+ \int\dd^d x'' |\bar{g}''|^{\frac{1}{2}} \bigg( \overline{\nabla}_{\alpha''} \bar{G}_{\mu\lambda''} \left( \frac{1}{2} h'' \bar{g}^{\alpha''\beta''} \bar{g}^{\lambda''\rho''}
- h^{\alpha''\beta''} \bar{g}^{\lambda''\rho''} \right) \overline{\nabla}_{\beta''} \bar{G}_{\rho''\nu'} \nnn
&\qquad \qquad + \frac{1}{2}
\overline{\nabla}^{\lambda''} \bar{G}_{\mu\sigma''}  h^{\alpha''\sigma''}
\overline{\nabla}_{\alpha''} \bar{G}_{\lambda''\nu'}
- \frac{1}{2}
\bar{G}_{\mu\sigma''}
h^{\alpha''\sigma''}
\overline{\nabla}_{\alpha''} \overline{\nabla}_{\nu'} \bar{G}_{x'',x'} \nnn
&\qquad \qquad + \frac{1}{2}
\overline{\nabla}_{\mu} \bar{G}_{x,x''} h^{\alpha''\lambda''}
\overline{\nabla}_{\alpha''} \bar{G}_{\lambda''\nu'}
- \frac{1}{2}
\bar{G}_{\mu\sigma''}
h^{\alpha''\lambda''}
\overline{\nabla}^{\sigma''} \overline{\nabla}_{\alpha''} \bar{G}_{\lambda''\nu'} \nnn
&\qquad \qquad - \frac{1}{2}
\overline{\nabla}_{\alpha''} \overline{\nabla}_{\mu} \bar{G}_{x,x''}
h^{\alpha''\sigma''}
\bar{G}_{\sigma''\nu'}
+ \frac{1}{2}
\overline{\nabla}_{\alpha''} \bar{G}_{\mu\lambda''}
h^{\alpha''\sigma''}
\overline{\nabla}^{\lambda''} \bar{G}_{\sigma''\nu'} \nnn
&\qquad \qquad - \frac{1}{2}
\overline{\nabla}^{\sigma''} \overline{\nabla}_{\alpha''} \bar{G}_{\mu\lambda''}
h^{\alpha''\lambda''} \bar{G}_{\sigma''\nu'}
+ \frac{1}{2}
\overline{\nabla}_{\alpha''} \bar{G}_{\mu\lambda''}
h^{\alpha''\lambda''}
\overline{\nabla}_{\nu'} \bar{G}_{x'',x'} \nnn
&\qquad \qquad + \bar{G}_{\mu\sigma''}
\left(
h^{\sigma''\rho''}
\bar{R}_{\rho''}^{\phantom{\rho''}\lambda''}
+ \bar{R}^{\sigma''}_{\phantom{\sigma''}\rho''} h^{\rho''\lambda''}
\right)
\bar{G}_{\lambda''\nu'} \nnn
&\qquad \qquad + \bar{G}_{\mu\lambda''} \left( (R|1)^{\lambda''\rho''} + \frac{1}{2} h'' \bar{R}^{\lambda''\rho''} \right) \bar{G}_{\rho''\nu'} \bigg) .
\end{align}}
\end{widetext}
In \eqref{GreensFunctionBornApproximation_Photon_gb+h}, we are again raising and lowering all indices with the background metric $\gb_{\mu\nu}$. Here and below, $(\mathcal{R}|n)$, $(R|n)_{\mu\nu}$ and $(R|n)_{\mu\nu \alpha\beta}$ are the portion of the respective geometric tensors (built out of $g_{\mu\nu} = \gb_{\mu\nu} + h_{\mu\nu}$) containing precisely $n$ powers of the perturbation $h_{\mu\nu}$.

When the background is Minkowski $\gb_{\mu\nu} = \eta_{\mu\nu}$ all the barred geometric tensors are identically zero. Like in the scalar case, we employ Cartesian coordinates and the spacetime translation symmetry property of $\bG_{x,x'}$ in \eqref{Minkowski_TranslationSymmetry} to massage \eqref{GreensFunctionBornApproximation_Photon_gb+h} into
\begin{widetext}
{\allowdisplaybreaks\begin{align}
\label{GreensFunctionBornApproximation_Photon_eta+h}
G_{\mu\nu'} &\approx \bar{G}_{x,x'} \eta_{\mu\nu} + \frac{1}{2} \bar{G}_{x,x'} \left( h_{\mu'\nu'} + h_{\mu\nu} \right) \\
            &\qquad \qquad + \int\dd^d x'' \bigg\{ \eta_{\mu\nu} \partial_{\alpha} \partial_{\beta'}
                    \bar{G}_{x,x''} \left( \frac{1}{2} h'' \eta^{\alpha''\beta''} - h^{\alpha''\beta''} \right) \bar{G}_{x'',x'} \nnn
&\qquad \qquad \qquad \qquad
            + \frac{1}{2} \left(\partial_\alpha - \partial_{\alpha'}\right)\left( \partial_{[\mu} + \partial_{[\mu'}\right)
                            \bar{G}_{x,x''} h^{\alpha''}_{\phantom{\alpha''} \nu'']} \bar{G}_{x'',x'}
            + \bar{G}_{x,x''} \left(R|1\right)_{\mu'' \nu''} \bar{G}_{x'',x'} \bigg\} . \nn
\end{align}}
\end{widetext}
This matches equation 2.23 of DeWitt and DeWitt \cite{DeWittDeWitt:1964}, up to a sign error, if we take into account both their $R_{\alpha\beta}$ and Green's function are negative of ours. (Their sign error\footnote{It is probably a typographic error, since DeWitt and DeWitt claimed their result satisfied \eqref{DivergenceOfPhotonG}.} is the following: the two terms on the line right before the last line (involving the Ricci tensor), should both carry a negative sign each, since they must have come from integrating by parts the term $-\delta[\sigma] h_{\sigma''\mu'',\nu''} \delta_,^{\sigma'}[\sigma']$.) As a consistency check of this result, one may perform a direct computation to show that the $G_{\mu\nu'}$ in \eqref{GreensFunctionBornApproximation_Photon_eta+h} satisfies \eqref{DivergenceOfPhotonG} to first order in $h_{\mu\nu}$.

{\bf Graviton} \quad Gravitation as encoded in the Einstein-Hilbert action
\begin{align}
\label{Einstein-Hilbert}
S_\text{EH} \equiv - \frac{1}{16\pi G_\text{N}} \int\dd^d x |g|^{\frac{1}{2}} \left( \mathcal{R} - 2\Lambda \right)
\end{align}
is a nonlinear theory. ($G_\text{N}$ is Newton's constant and $\Lambda$ is the cosmological constant.) One can insert the metric $g_{\mu\nu} + \sqrt{32\pi G_\text{N}} \gamma_{\mu\nu}$ into the Einstein-Hilbert action \eqref{Einstein-Hilbert} and find a resulting infinite series in $\gamma_{\mu\nu}$. The quadratic piece, which will determine for us the Green's function of the graviton, is
{\allowdisplaybreaks\begin{align}
\label{GravitonAction}
&S_\gamma[g]
= \frac{1}{2} \int \dd^d x |g|^{\frac{1}{2}} \bigg( \nabla^\mu \gamma^{\beta\nu} \nabla_{\mu} \gamma_{\beta\nu}
- \frac{1}{2} \nabla^\sigma \gamma \nabla_\sigma \gamma \nnn
&\qquad - 2 R_{\nu\lambda\beta\mu} \gamma^{\lambda\mu} \gamma^{\beta\nu}
- 2 \gamma^{\beta\sigma} \gamma_\sigma^{\phantom{\sigma}\nu} R_{\beta\nu}
+ 2 \gamma \ \gamma^{\beta\nu} R_{\beta\nu} \nnn
&\qquad + \left( \gamma_{\sigma\rho} \gamma^{\sigma\rho} - \frac{1}{2} \gamma^2 \right) \left( \mathcal{R} - 2\Lambda
\right) \bigg) ,
\end{align}}
where we have chosen the de Donder gauge $\nabla^\mu \gamma_{\mu\nu} = \frac{1}{2} \nabla_\nu \gamma$, with $\gamma \equiv g^{\mu\nu} \gamma_{\mu\nu}$. (The geometric tensors in \eqref{GravitonAction}, such as $R_{\nu\lambda\beta\mu}$, are built out of $g_{\mu\nu}$.) From \eqref{GravitonAction} and following the preceding analysis for the scalar and photon, we may write down the integral equation involving the graviton Green's functions
\begin{widetext}
{\allowdisplaybreaks\begin{align}
\label{GreensFunctionIntegralEquation_Graviton}
&G_{\delta\epsilon \rho'\sigma'} - \bar{G}_{\delta\epsilon \rho'\sigma'} \nnn
&= \int \dd^d x'' |g''|^{\frac{1}{2}} \bigg(
\nabla_{\tau''} \bar{G}_{\delta\epsilon \alpha''\beta''} g^{\tau''\kappa''} \left( g^{\alpha''\mu''} g^{\beta''\nu''} - \frac{1}{2} g^{\alpha''\beta''} g^{\mu''\nu''} \right) \nabla_{\kappa''} G_{\mu''\nu'' \rho'\sigma'} \nnn
&\qquad + \bar{G}_{\delta\epsilon \alpha''\beta''} \bigg( \left( g^{\alpha''\mu''} g^{\beta''\nu''} - \frac{1}{2}  g^{\alpha''\beta''} g^{\mu''\nu''} \right)
    \left( \mathcal{R} - 2\Lambda \right) - 2 R^{\mu'' \alpha'' \nu'' \beta''} \nnn
&\qquad \qquad - R^{\beta''\nu''} g^{\alpha''\mu''} -  R^{\alpha''\nu''} g^{\beta''\mu''}
+ R^{\mu''\nu''} g^{\alpha''\beta''} + R^{\alpha''\beta''} g^{\mu''\nu''} \bigg) G_{\mu''\nu'' \rho'\sigma'} \bigg) \nnn
&\qquad -\int \dd^d x'' |\gb''|^{\frac{1}{2}} \bigg(
\Db_{\tau''} G_{\delta\epsilon \alpha''\beta''} \bar{g}^{\tau''\kappa''} \left( \gb^{\alpha''\mu''} \gb^{\beta''\nu''} - \frac{1}{2} \gb^{\alpha''\beta''} \gb^{\mu''\nu''} \right) \Db_{\kappa''} \bar{G}_{\mu''\nu'' \rho'\sigma'}  \nnn
&\qquad + G_{\delta\epsilon \alpha''\beta''} \bigg(
\left( \gb^{\alpha''\mu''} \gb^{\beta''\nu''} - \frac{1}{2}  \gb^{\alpha''\beta''} \gb^{\mu''\nu''} \right)
    \left( \bar{\mathcal{R}} - 2\Lambda \right) - 2 \bar{R}^{\mu'' \alpha'' \nu'' \beta''} \nnn
&\qquad \qquad - \bar{R}^{\beta''\nu''} \gb^{\alpha''\mu''} -  \bar{R}^{\alpha''\nu''} \gb^{\beta''\mu''}
+ \bar{R}^{\mu''\nu''} \gb^{\alpha''\beta''} + \bar{R}^{\alpha''\beta''} \gb^{\mu''\nu''} \bigg) \bar{G}_{\mu''\nu'' \rho'\sigma'} \bigg) .
\end{align}}
\end{widetext}

\emph{First Born Approximation} \quad Because of the number of terms and the plethora of indices in \eqref{GreensFunctionIntegralEquation_Graviton}, the perturbation theory about a generic background $\gb_{\mu\nu}$ and arbitrary dimensions $d$ is best left for a computer algebra system to handle. We shall be content with the case of 4 dimensional perturbed Minkowski spacetime, and also set the cosmological constant to zero for now. To $\mathcal{O}[h]$, we replace in \eqref{GreensFunctionIntegralEquation_Graviton} all the $G_{\delta\epsilon \alpha'\beta'}$ occurring under the integral sign with $\bar{P}_{\delta\epsilon \alpha\beta} \bG_{x,x'}$ (see \eqref{GreensFunctionMinkowski_Graviton}) and expand all quantities about Minkowski spacetime up to first order in perturbations. Let us employ Cartesian coordinates, raise and lower indices with $\eta_{\mu\nu}$, and integrate-by-parts the derivatives acting on $h_{\mu\nu}$ occurring within the Christoffel symbols,
\begin{align}
\label{IBPIdentity}
&\int \dd^d x'' \bG_{x,x''} \partial_{\mu''} h_{\alpha''\beta''} \bG_{x'',x'} \\
&\qquad = (\partial_\mu + \partial_{\mu'})\int \dd^d x'' \bG_{x,x''} h_{\alpha''\beta''} \bG_{x'',x'}, \nn
\end{align}
where we have invoked \eqref{Minkowski_TranslationSymmetry}. It helps to exploit the symmetries of the Riemann tensor indices ($R_{\mu\nu \alpha\beta} = R_{\alpha\beta \mu\nu} = -R_{\nu\mu \alpha\beta} = -R_{\mu\nu \beta\alpha}$), those of $\bar{P}$ recorded in \eqref{PbarSymmetries}, and to recognize that, in $d=4$ dimensions,
\begin{align}
\bar{P}^{\alpha\beta \mu\nu} \bar{P}_{\mu\nu \rho\sigma} = \frac{1}{2} \delta^\alpha_{\{\rho} \delta^\beta_{\sigma\}} .
\end{align}
For reasons to be apparent in the next section, we shall re-express all the $\partial_\tau \partial^{\tau'}$ as
\begin{align}
\partial_\tau \partial^{\tau'} &= \frac{1}{2}(\partial_\tau + \partial_{\tau'})(\partial^\tau + \partial^{\tau'}) - \frac{1}{2} \partial_\tau \partial^\tau - \frac{1}{2} \partial_{\tau'} \partial^{\tau'} \nnn
&\equiv \frac{1}{2}(\partial+\partial')^2 - \frac{1}{2}\partial^2 - \frac{1}{2}\partial'^2 \nn
\end{align}
followed by using the Minkowski version of \eqref{GreensFunctionPDE_Scalar}, namely
\begin{align}
\partial^2 \bG_{x,x'} = \partial'^2 \bG_{x,x'} = \delta^d[x-x'] .
\end{align}
We then arrive at
\begin{widetext}
{\allowdisplaybreaks\begin{align}
\label{GreensFunctionBornApproximation_Graviton_eta+h}
G_{\delta\epsilon \rho'\sigma} &\approx
\bar{G}_{x,x'} \bigg(
    \bar{P}_{\delta\epsilon \rho\sigma} +
    \frac{1}{4}
    \left(
        \eta_{\rho\delta} (h_{\epsilon\sigma}+h_{\epsilon'\sigma'})
        + \eta_{\sigma\delta} (h_{\epsilon\rho}+h_{\epsilon'\rho'})
        + \eta_{\rho\epsilon} (h_{\delta\sigma}+h_{\delta'\sigma'})
        + \eta_{\sigma\epsilon} (h_{\delta\rho}+h_{\delta'\rho'})
    \right) \nnn
&\qquad \qquad \qquad
    - \frac{1}{2} \eta_{\rho\sigma} h_{\delta\epsilon} - \frac{1}{2} \eta_{\delta\epsilon} h_{\rho'\sigma'} \bigg) \nnn
&+ \int \dd^4 x''\bigg\{
\bar{P}_{\delta\epsilon \rho\sigma}
    \partial_{\alpha} \partial_{\beta'} \bar{G}_{x,x''} \left(\frac{1}{2} h'' \eta^{\alpha\beta} - h^{\alpha''\beta''} \right) \bar{G}_{x'',x'} \nnn
&\qquad \qquad
+ \frac{1}{4} \left( \partial^\tau - \partial^{\tau'} \right) \bigg(
    (\partial_{[\epsilon} + \partial_{[\epsilon'}) \bar{G}_{x,x''} h_{\rho'']\tau''} \bar{G}_{x'',x'} \eta_{\sigma\delta}
    + (\partial_{[\epsilon} + \partial_{[\epsilon'}) \bar{G}_{x,x''} h_{\sigma'']\tau''} \bar{G}_{x'',x'} \eta_{\rho\delta} \nnn
&\qquad \qquad \qquad \qquad \qquad \qquad
    + (\partial_{[\delta} + \partial_{[\delta'}) \bar{G}_{x,x''} h_{\rho'']\tau''} \bar{G}_{x'',x'} \eta_{\sigma\epsilon}
    + (\partial_{[\delta} + \partial_{[\delta'}) \bar{G}_{x,x''} h_{\sigma'']\tau''} \bar{G}_{x'',x'} \eta_{\rho\epsilon} \bigg) \nnn
&\qquad \qquad
+ \bar{G}_{x,x''} \big(
    \bar{P}_{\delta\epsilon \rho\sigma} (\mathcal{R}|1)
    + \eta_{\epsilon\delta} (R|1)_{\rho''\sigma''} + \eta_{\rho\sigma} (R|1)_{\delta''\epsilon''} \nnn
    &\qquad \qquad \qquad \qquad
    - \frac{1}{2} \eta_{\rho\{\delta} (R|1)_{\epsilon''\}\sigma''} - \frac{1}{2} \eta_{\sigma\{\delta} (R|1)_{\epsilon''\}\rho''}
    + (R|1)_{\rho''\{\delta''\epsilon''\}\sigma''}
\big) \bar{G}_{x'',x'} \bigg\} .
\end{align}}
\end{widetext}
{\bf One scattering approximation} \quad Let us examine \eqref{GreensFunctionBornApproximation_Scalar_eta}, \eqref{GreensFunctionBornApproximation_Photon_eta+h} and \eqref{GreensFunctionBornApproximation_Graviton_eta+h}. The terms that do not involve any integrals can be viewed as the propagation of null signals, modulated by the metric perturbations multiplying the $\bG_{x,x'}$. The terms involving integrals, go schematically as $\partial_x \partial_{x'} \int \dd^4 x'' \bG_{x,x''} h[x''] \bG_{x'',x'}$. Due to the causal structure of the $\bG$s, this can be interpreted as a scattering process. The $\bG_{x'',x'}$ tells us our massless field begins at the source $x'$ and travels along a null ray to $x''$; the $h[x'']$ says it then scatters off the metric perturbations (and its derivatives) at $x''$; and the $\bG_{x,x''}$ informs us that it then propagates along a null path from $x''$ to reach the observer at $x$. The full (scattered) signal consists of integrating over all the $x''$ from which the signal can scatter off. This is the perturbative picture for the origin of tails of massless fields in weakly curved spacetime.\footnote{We are being slightly inaccurate here, in that some of the $\partial_x \partial_{x'} \int \dd^4 x'' \bG_{x,x''} h[x''] \bG_{x'',x'}$ terms also contribute to null propagation, as we will see in the next section. But we want to introduce this scattering picture here, because it is easier to see it from \eqref{GreensFunctionBornApproximation_Scalar_eta}, \eqref{GreensFunctionBornApproximation_Photon_eta+h} and \eqref{GreensFunctionBornApproximation_Graviton_eta+h}, written in terms of the Minkowski Green's function $\bG$s, than from \eqref{GreensFunctionBornApproximation_Scalar_eta_Decomposed}, \eqref{GreensFunctionBornApproximation_Photon_eta_Decomposed}, and \eqref{GreensFunctionBornApproximation_Graviton_eta_Decomposed} below, which are expressed in terms of $\Theta[\sigma_{x,x'}]$ and the $\widehat{\mathcal{I}}$-integrals in \eqref{MasterMatrix_II}.} From this heuristic point of view, we can already anticipate that high order perturbation theory will involve more than one scattering events contributing to the tail effect. This scattering picture may also help us estimate its size without detailed calculations, and deserves some contemplation.

\section{$\delta[\sigma]$ and $\Theta[\sigma]$ decomposition in 4 dimensional perturbed Minkowski}

In this section we will restrict ourselves to 4 dimensions and analyze further the first order results for the scalar \eqref{GreensFunctionBornApproximation_Scalar_eta}, photon \eqref{GreensFunctionBornApproximation_Photon_eta+h} and graviton \eqref{GreensFunctionBornApproximation_Graviton_eta+h} Green's functions we have obtained in perturbed Minkowski spacetime, and show that to $\mathcal{O}[h]$, concrete results for the Green's functions can be gotten once a single matrix of integrals (involving $h_{\alpha\beta}$) can be performed. We will also decompose these scalar, photon and graviton Green's functions into their null cone and tail pieces. As a consistency check of our Born approximation, we show that their null cone pieces matches the Hadamard form described by equations \eqref{NullConeFunction_Scalar}, \eqref{NullConeFunction_Photon} and \eqref{NullConeFunction_Graviton}; this generalizes the analysis carried out in Pfenning and Poisson \cite{PfenningPoisson:2000zf} to the case of arbitrary perturbations $h_{\mu\nu}$.

In the scalar \eqref{GreensFunctionBornApproximation_Scalar_eta}, photon \eqref{GreensFunctionBornApproximation_Photon_eta+h} and graviton \eqref{GreensFunctionBornApproximation_Graviton_eta+h} Green's functions results, we have to deal with derivatives (with respect to $x$ or $x'$) acting on the following matrix integral
\begin{align}
\label{MasterIntegral}
\frac{1}{4\pi} \mathcal{I}_{\alpha\beta} \equiv \int \dd^4 x'' \bG_{x,x''} h_{\alpha''\beta''} \bG_{x'',x'}
\end{align}
with the $\bG_{x,x'}$ from \eqref{GreensFunctionMinkowski_Scalar}. Because of \eqref{IBPIdentity}, even the geometric tensor terms can be expressed as sum of derivatives with respect to $x$ or $x'$ acting on \eqref{MasterIntegral}. For example,
\begin{align}
\label{G_RicciTensor_G}
&\int \dd^4 x'' \bG_{x,x''} (\mathcal{R}|1) \bG_{x'',x'} \\
&\qquad = \partial_{\alpha^+} \partial_{\beta^+} \int \dd^4 x'' \bG_{x,x''} (h^{\alpha''\beta''} - \eta^{\alpha\beta} h'') \bG_{x'',x'} \nn
\end{align}
where $h''$ is the trace of $h_{\mu''\nu''}$ and
\begin{align}
\partial_{\alpha^+} \equiv \partial_\alpha + \partial_{\alpha'} .
\end{align}
In appendix \eqref{Appendix_TheMatrix} we show that $\mathcal{I}_{\alpha\beta}$ involves the integral of $h_{\alpha\beta}$ (but in Euclidean 3-space) over the surface generated by rotating the ellipse with foci at $\vec{x}$ and $\vec{x}'$ and semi-major axis $(t-t')/2$, about the line joining $\vec{x}$ and $\vec{x}'$. (This is the dashed oval in Fig. \eqref{TwoLightConesIntersectFigure}.)
\begin{align}
\label{MasterMatrix_I}
\mathcal{I}_{\alpha\beta}[x,x'] \equiv \Theta[t-t']\Theta[\bar{\sigma}_{x,x'}] \widehat{\mathcal{I}}_{\alpha\beta}[x,x']
\end{align}
with
\begin{align}
\label{MasterMatrix_II}
&\widehat{\mathcal{I}}_{\alpha\beta}[x,x'] \\
&= \frac{1}{2} \int_{\mathbb{S}^2} \frac{\dd \Omega}{4\pi}
h_{\alpha''\beta''}
\left[
\frac{t+t'}{2} + \frac{|\vec{\Delta}|}{2} \cos \theta,
\frac{\vec{x} + \vec{x}'}{2} + \vec{x}''
\right] . \nn
\end{align}
The infinitesimal solid angle is $\dd \Omega = \dd \cos\theta \dd \phi$, and the Cartesian components of $\vec{x}''$ are given by
\begin{align}
\label{x''}
\vec{x}'' \equiv
\left(
\sqrt{\frac{\bar{\sigma}_{x,x'}}{2}} \sin\theta \cos\phi,
\sqrt{\frac{\bar{\sigma}_{x,x'}}{2}} \sin\theta \sin\phi,
\frac{\Delta^0}{2} \cos\theta
\right) .
\end{align}
To separate the light cone versus tail pieces of the Green's functions, we now carry out the necessary derivatives on \eqref{MasterMatrix_I} as they occur in \eqref{GreensFunctionBornApproximation_Scalar_eta}, \eqref{GreensFunctionBornApproximation_Photon_eta+h} and \eqref{GreensFunctionBornApproximation_Graviton_eta+h}. There is no need to differentiate the $\Theta[t-t']$, because that would give us $\delta[t-t']$ and its derivatives. Since this would be multiplied by either $\Theta[\bar{\sigma}_{x,x'}]$ or possibly $\delta[\bar{\sigma}_{x,x'}]$, $\delta'[\bar{\sigma}_{x,x'}]$, etc., while $\bar{\sigma}_{x,x'} \to -\vec{\Delta}^2/2 < 0$, these $\delta,\delta',\dots$ terms can never be non-zero when $t=t'$. Schematically, therefore, the derivatives now read $\Theta[t-t'] \partial \partial (\Theta[\bar{\sigma}] \widehat{\mathcal{I}})$ (where the two derivatives are both with respect to either $x$ or $x'$ or one each), which in turn would yield two types of terms. One is the tail term, proportional to $\Theta[\bar{\sigma}] \partial \partial  \widehat{\mathcal{I}}$ and the other the null cone ones, proportional to either $\delta[\bar{\sigma}]\partial \bar{\sigma} \partial \widehat{\mathcal{I}}$, $\delta[\bar{\sigma}] \partial \partial \bar{\sigma} \widehat{\mathcal{I}}$, or $\delta'[\bar{\sigma}] \partial \bar{\sigma} \partial \bar{\sigma} \widehat{\mathcal{I}}$. Following that, we would impose the constraint $\bar{\sigma}_{x,x'} = 0$ on the coefficients of the $\delta[\bar{\sigma}]$ and $\delta'[\bar{\sigma}]$ terms. This requires that we develop a power series in $\bar{\sigma}_{x,x'}$ of $\widehat{\mathcal{I}}_{\alpha\beta}$. Since there is at most one derivative acting on $\widehat{\mathcal{I}}$, however, we only need to do so up to linear order. (Higher order terms would automatically vanish once we put $\bar{\sigma} = 0$.) In appendix \eqref{Appendix_TheMatrix} we find
\begin{align}
\label{MasterMatrix_NullConeExpansion}
\widehat{\mathcal{I}}_{\alpha\beta}
    &= \widehat{\mathcal{I}}^{(0)}_{\alpha\beta} + \bar{\sigma}_{x,x'} \widehat{\mathcal{I}}^{(1)}_{\alpha\beta} + \dots \\
    &= \left( 1 - \frac{\bar{\sigma}_{x,x'}}{2} \eta^{\mu\nu} \partial_\mu \partial_{\nu'} \right) \frac{1}{2} \int_0^1 h_{\alpha''\beta''}[\bar{\xi}] \dd \lambda \nn
+ \dots .
\end{align}
The $\widehat{\mathcal{I}}^{(0)}_{\alpha\beta} = \frac{1}{2} \int_0^1 h_{\alpha''\beta''}[\bar{\xi}] \dd \lambda$ has already been quoted previously in \eqref{I0}.

{\bf Scalar} \quad By pulling out one factor of $(4\pi)^{-1}$ from one of the $\bG$s (see \eqref{GreensFunctionMinkowski_Scalar}), our result for the massless scalar Green's function in \eqref{GreensFunctionBornApproximation_Scalar_eta} can be written as
\begin{align}
&G_{x,x'} \approx \frac{\Theta[t-t']}{4\pi} \bigg( \delta[\bar{\sigma}_{x,x'}] \nonumber\\
&\qquad + \partial_\alpha \partial_{\beta'} \left( \left\{ \frac{1}{2} \eta^{\alpha\beta} \widehat{\mathcal{I}} - \widehat{\mathcal{I}}^{\alpha\beta} \right\} \Theta[\bar{\sigma}_{x,x'}] \right) \bigg) ,
\end{align}
with $\widehat{\mathcal{I}} \equiv \widehat{\mathcal{I}}^{\rho\kappa} \eta_{\rho\kappa}$. Carrying out the derivatives using \eqref{Derivatives} would give us, amongst other terms, the following $\delta'$ terms:
\begin{align}
- \delta'\left[ \bar{\sigma}_{x,x'} \right] \left( \bar{\sigma}_{x,x'} \widehat{\mathcal{I}} - \Delta^\alpha \Delta^\beta \widehat{\mathcal{I}}_{\alpha\beta} \right) .
\end{align}
The first term is $\delta\left[ \bar{\sigma}_{x,x'} \right] \widehat{\mathcal{I}}$ if we employ the identity $z\delta'[z]= -\delta[z]$. The second term can be considered the $\mathcal{O}[h]$ term of $\delta[\bar{\sigma} + \Delta^\alpha \Delta^\beta \widehat{\mathcal{I}}_{\alpha\beta}] = \delta[\bar{\sigma}] + \delta'[\bar{\sigma}] \Delta^\alpha \Delta^\beta \widehat{\mathcal{I}}_{\alpha\beta} + \dots$.

Moreover, invoking \eqref{DerivativeLambda} and the chain rule also informs us that one of the terms multiplying $\delta[\bar{\sigma}_{x,x'}]$ is
\begin{align}
-\frac{1}{2} \Delta^\kappa (\partial_\kappa - \partial_{\kappa'}) \widehat{\mathcal{I}}^{(0)}_{\alpha\beta}
&= -\frac{1}{4} h_{\alpha\beta} -\frac{1}{4} h_{\alpha'\beta'} + \widehat{\mathcal{I}}^{(0)}_{\alpha\beta} .
\end{align}

Altogether, the Born approximation, $\mathcal{O}[h]$-accurate answer, for the massless scalar Green's function may now be decomposed into its null cone and tail pieces as
\begin{widetext}
{\allowdisplaybreaks\begin{align}
\label{GreensFunctionBornApproximation_Scalar_eta_Decomposed}
G_{x,x'} &\approx \frac{\Theta[t-t']}{4\pi} \bigg\{ \delta\left[ \bar{\sigma}_{x,x'} + \Delta^\alpha \Delta^\beta \widehat{\mathcal{I}}_{\alpha\beta}^{(0)} \right]
\left( 1
- \frac{1}{4} h -\frac{1}{4} h' + \widehat{\mathcal{I}}^{(0)}
+ \Delta^\alpha \left( \partial^\beta - \partial^{\beta'} \right) \widehat{\mathcal{I}}^{(0)}_{\alpha\beta}
- \frac{1}{2} \Delta^\alpha \Delta^\beta \partial^{\mu'} \partial_\mu \widehat{\mathcal{I}}^{(0)}_{\alpha\beta}
\right) \nnn
&\qquad + \Theta\left[\bar{\sigma}_{x,x'} + \Delta^\alpha \Delta^\beta \widehat{\mathcal{I}}_{\alpha\beta}^{(0)}\right] \left( \frac{1}{2} \partial^\mu \partial_{\mu'} \widehat{\mathcal{I}} - \partial_\rho \partial_{\kappa'} \widehat{\mathcal{I}}^{\rho\kappa} \right) \bigg\}, \qquad
    h \equiv \eta^{\alpha\beta} h_{\alpha\beta}; \ h' \equiv \eta^{\alpha\beta} h_{\alpha'\beta'} .
\end{align}}
\end{widetext}
As already advertised earlier, comparison with \eqref{WorldFunction_PerturbedMinkowski} and \eqref{vanVleckDeterminant_PerturbedMinkowski} tells us the null cone portion of our massless scalar Green's function is indeed consistent with the Hadamard form in \eqref{HadamardForm_Scalar} and \eqref{NullConeFunction_Scalar}.

{\bf Photon and Graviton} \quad For the photon $G_{\mu\nu'}$ \eqref{GreensFunctionBornApproximation_Photon_eta+h} and graviton $G_{\delta\epsilon \rho'\sigma'}$ \eqref{GreensFunctionBornApproximation_Graviton_eta+h} Green's functions, we first observe that they contain respectively $\eta_{\mu\nu}$ and $\bar{P}_{\delta\epsilon \rho'\sigma'}$ multiplied by \eqref{GreensFunctionBornApproximation_Scalar_eta_Decomposed}, the massless scalar $G_{x,x'}$. (Specifically, first term on the first line, and the second line of \eqref{GreensFunctionBornApproximation_Photon_eta+h} for the photon; and first term on the first line, and third line of \eqref{GreensFunctionBornApproximation_Graviton_eta+h} for the graviton.) The light cone portions of these terms therefore contain the first order van Vleck determinant. For the rest of the integral terms, we first make the observation that $\partial_\mu + \partial_{\mu'}$ acting on a function whose argument is the difference $x-x'$, is identically zero. The immediate corollary is that all the geometric terms, via \eqref{IBPIdentity}, do not contribute to the null cone piece of the photon and graviton Green's function because the derivatives acting on the $\Theta[\bar{\sigma}]$ leads to zero. The remaining terms containing derivatives take the form
{\allowdisplaybreaks\begin{align}
\label{ParallelPropagator_PerturbedMinkowski_MissingPiece}
&\frac{1}{2} \left(\partial^\tau - \partial^{\tau'}\right) \left( \partial_{[\mu} + \partial_{[\mu'} \right) \left( \Theta[\bar{\sigma}_{x,x'}] \widehat{\mathcal{I}}_{\nu]\tau} \right) \nnn
&= \delta[\bar{\sigma}_{x,x'}] \frac{\Delta^\rho}{2} \int_0^1 \partial_{[\mu''} h_{\nu'']\rho''}[\bar{\xi}] \dd \lambda \\
&\qquad \qquad + \Theta[\bar{\sigma}_{x,x'}] \frac{1}{2} \left(\partial^\tau - \partial^{\tau'}\right) \left( \partial_{[\mu} + \partial_{[\mu'} \right) \widehat{\mathcal{I}}_{\nu]\tau} , \nn
\end{align}}
where we have utilized \eqref{Derivatives} and the chain rule. Recalling \eqref{ParallelPropagator_PerturbedMinkowski} tells us the $\delta[\bar{\sigma}]$ terms on the right side of \eqref{ParallelPropagator_PerturbedMinkowski_MissingPiece}, when added to the non-integral $\mathcal{O}[h]$ ones already multiplying $\bar{G}_{x,x'}$ -- i.e., the first line of \eqref{GreensFunctionBornApproximation_Photon_eta+h} and first two lines of \eqref{GreensFunctionBornApproximation_Graviton_eta+h} -- would give us the necessary first order parallel propagators to once again ensure consistency with the Hadamard form in \eqref{HadamardForm_Photon}, \eqref{NullConeFunction_Photon}, \eqref{HadamardForm_Graviton} and \eqref{NullConeFunction_Graviton}. That is, we may now use the expressions for $\sigma_{x,x'}$ \eqref{WorldFunction_PerturbedMinkowski}, $\sqrt{\Delta_{x,x'}}$ \eqref{vanVleckDeterminant_PerturbedMinkowski} and $g_{\mu\nu'}$ \eqref{ParallelPropagator_PerturbedMinkowski} and decompose the photon and graviton Green's function into their null cone and tail pieces. To first order in $h_{\mu\nu}$,
\begin{widetext}
\begin{align}
\label{GreensFunctionBornApproximation_Photon_eta_Decomposed}
&G_{\mu\nu'} \approx \frac{\Theta[t-t']}{4\pi} \bigg\{
g_{\mu\nu'} \sqrt{\Delta_{x,x'}} \delta\left[\sigma_{x,x'}\right] \\
&\qquad \qquad \qquad \qquad + \Theta\left[\sigma_{x,x'}\right] \left(
\eta_{\mu\nu} \left(\frac{1}{2} \partial^\alpha \partial_{\alpha'} \widehat{\mathcal{I}} - \partial_\alpha \partial_{\beta'} \widehat{\mathcal{I}}^{\alpha\beta}\right)
    + \frac{1}{2} \left( \partial_\alpha - \partial_{\alpha'} \right) \left( \partial_{[\mu} + \partial_{[\mu'} \right) \widehat{\mathcal{I}}^\alpha_{\phantom{\alpha}\nu]}
+ \widehat{(\mathcal{R}|1)}_{\mu\nu}
\right)
\bigg\} , \nn
\end{align}
{\allowdisplaybreaks\begin{align}
\label{GreensFunctionBornApproximation_Graviton_eta_Decomposed}
&G_{\delta\epsilon \rho'\sigma'}
\approx \frac{\Theta[t-t']}{4\pi} \bigg\{ P_{\delta\epsilon \rho'\sigma'} \sqrt{\Delta_{x,x'}} \delta\left[ \sigma_{x,x'} \right] \nnn
&\qquad \qquad \qquad \qquad + \Theta\left[ \sigma_{x,x'} \right] \bigg(
\bar{P}_{\delta\epsilon \rho\sigma}
    \left(\frac{1}{2} \partial^\alpha \partial_{\alpha'} \widehat{\mathcal{I}} - \partial_\alpha \partial_{\beta'} \widehat{\mathcal{I}}^{\alpha\beta}\right) \nnn
&\qquad \qquad \qquad \qquad \qquad \qquad
+ \frac{1}{4} \left( \partial^\tau - \partial^{\tau'} \right)
\bigg(
    (\partial_{[\epsilon} + \partial_{[\epsilon'}) \widehat{\mathcal{I}}_{\rho'']\tau''} \eta_{\sigma\delta}
    + (\partial_{[\epsilon} + \partial_{[\epsilon'}) \widehat{\mathcal{I}}_{\sigma'']\tau''} \eta_{\rho\delta} \nnn
&\qquad \qquad \qquad \qquad \qquad \qquad \qquad \qquad
    + (\partial_{[\delta} + \partial_{[\delta'}) \widehat{\mathcal{I}}_{\rho'']\tau''} \eta_{\sigma\epsilon}
    + (\partial_{[\delta} + \partial_{[\delta'}) \widehat{\mathcal{I}}_{\sigma'']\tau''} \eta_{\rho\epsilon} \bigg) \nnn
&\qquad \qquad \qquad \qquad
    + \bar{P}_{\delta\epsilon \rho\sigma} \widehat{(\mathcal{R}|1)}
    + \eta_{\epsilon\delta} \widehat{(R|1)}_{\rho''\sigma''} + \eta_{\rho\sigma} \widehat{(R|1)}_{\delta''\epsilon''}
    - \frac{1}{2} \eta_{\rho\{\delta} \widehat{(R|1)}_{\epsilon''\}\sigma''} - \frac{1}{2} \eta_{\sigma\{\delta} \widehat{(R|1)}_{\epsilon''\}\rho''}  \nnn
&\qquad \qquad \qquad \qquad \qquad \qquad
    + \widehat{(R|1)}_{\rho''\{\delta''\epsilon''\}\sigma''}
\bigg)
\bigg\} .
\end{align}}
\end{widetext}
The geometric terms $\widehat{(R|1)}_{\alpha\beta\mu\nu}$, $\widehat{(R|1)}_{\mu\nu}$, and $\widehat{(\mathcal{R}|1)}$ in \eqref{GreensFunctionBornApproximation_Photon_eta_Decomposed} and \eqref{GreensFunctionBornApproximation_Graviton_eta_Decomposed} can be obtained by taking the corresponding linearized tensors in terms of the perturbation $h_{\alpha\beta}$, and replacing all the $h_{\alpha\beta}$ with $\widehat{\mathcal{I}}_{\alpha\beta}$ and all derivatives $\partial_{\alpha''}$ with $\partial_{\alpha^+}$.
{\allowdisplaybreaks\begin{align}
\widehat{(R|1)}_{\alpha\beta\mu\nu} &\equiv
    \frac{1}{2} \left( \partial_{\beta^+} \partial_{[\mu^+} \widehat{\mathcal{I}}_{\nu]\alpha} - \partial_{\alpha^+} \partial_{[\mu^+} \widehat{\mathcal{I}}_{\nu]\beta} \right) \\
\widehat{(R|1)}_{\beta\nu} &\equiv
    \frac{1}{2} \bigg( \partial_{\mu^+} \partial_{\{\beta^+} \widehat{\mathcal{I}}_{\nu\}}^{\phantom{\mu}\mu} \nonumber\\
    &\qquad - \partial_{\beta^+} \partial_{\nu^+} \widehat{\mathcal{I}} - \eta^{\alpha\mu} \partial_{\alpha^+} \partial_{\mu^+} \widehat{\mathcal{I}}_{\beta\nu} \bigg) \\
\widehat{(\mathcal{R}|1)} &\equiv
    \partial_{\alpha^+} \partial_{\beta^+} \left( \widehat{\mathcal{I}}^{\alpha\beta} - \eta^{\alpha\beta}  \widehat{\mathcal{I}} \right) ,
\end{align}}
with $\partial_{\alpha^+} \equiv \partial_\alpha + \partial_{\alpha'}$. Even though these terms involving geometric curvature are best evaluated by differentiating $\widehat{\mathcal{I}}_{\alpha\beta}$, it is necessary to record here their analogs to \eqref{MasterMatrix_II}. For instance, if $(R|1)_{\alpha\beta}$ is the linearized Ricci tensor, we have
\begin{align}
\label{MasterMatrix_Ricci_I}
&\frac{1}{4\pi} \Theta[t-t']\Theta[\bar{\sigma}_{x,x'}] \widehat{(R|1)}_{\alpha\beta} \nonumber\\
& \qquad \qquad =\int \dd^4 x'' \bG_{x,x''} (R|1)_{\alpha\beta} \bG_{x'',x'},
\end{align}
where
\begin{align}
\label{MasterMatrix_Ricci_II}
&\widehat{(R|1)}_{\alpha\beta}[x,x'] \\
&= \frac{1}{2} \int_{\mathbb{S}^2} \frac{\dd \Omega}{4\pi}
(R|1)_{\alpha''\beta''}
\left[
\frac{t+t'}{2} + \frac{|\vec{\Delta}|}{2} \cos \theta,
\frac{\vec{x} + \vec{x}'}{2} + \vec{x}''
\right] . \nn
\end{align}
At the first Born approximation, therefore, we see that a concrete expression from the perturbative solution of the scalar \eqref{GreensFunctionBornApproximation_Scalar_eta_Decomposed}, photon \eqref{GreensFunctionBornApproximation_Photon_eta_Decomposed}, and graviton \eqref{GreensFunctionBornApproximation_Graviton_eta_Decomposed} can be obtained once the matrix integral $\widehat{\mathcal{I}}_{\alpha\beta}$ in \eqref{MasterMatrix_II} is evaluated. We also note that, suppose $\widehat{\mathcal{I}}_{\alpha\beta}$ in \eqref{MasterMatrix_II} has been evaluated; then at least when $h_{\alpha\beta}$ is time-independent (space-independent), there is no need to perform the line integral $\widehat{\mathcal{I}}^{(0)}_{\alpha\beta}$ in \eqref{MasterMatrix_NullConeExpansion}; rather, $\widehat{\mathcal{I}}^{(0)}_{\alpha\beta}$ is gotten by replacing $t-t' \to |\vec{x}-\vec{x}'|$ ($|\vec{x}-\vec{x}'| \to t-t'$). In such cases, the Born series method advocated here allows one to read off, as a byproduct of a single coherent calculation, the world function and van Vleck determinant from, respectively, the argument and coefficient of the $\delta$-function in the massless scalar Green's function; while the parallel propagator can be read off the coefficient of the $\delta$-function in the Lorenz gauge photon Green's function.

{\bf Gauge dependence} \quad The skeptic may wonder if the gauge dependence of the vector potential could render the tail piece of the photon Green's function in \eqref{GreensFunctionBornApproximation_Photon_eta_Decomposed} un-physical. To that end, we note that, for fixed $x'$, the only pure gradient tail term in \eqref{GreensFunctionBornApproximation_Photon_eta_Decomposed} is $(1/2)\partial_\mu (\partial_\alpha - \partial_{\alpha'}) \widehat{\mathcal{I}}^\alpha_{\phantom{\alpha}\nu}$. Hence, the rest of the tail terms do not have zero curl -- the corresponding electromagnetic fields are non-zero. This provides strong theoretical evidence that the wake effect is present for photons propagating in perturbed Minkowski, and by conformal symmetry, in our universe too.

{\bf Geometry and tails} \quad Let us notice that it was all the differentiation that took place in our work on the perturbative solution of the Green's functions, which can be traced to the $\Box$ operator, that gave us both the terms in the arguments and coefficients of the $\delta$-functions in the scalar, photon, and graviton Green's function. In turn, we have identified them as various terms in the world function, the van Vleck determinant and the parallel propagator (in their perturbative guises). This re-affirms our assertion earlier that it is the differential operator $\Box$ that is solely responsible for the behavior of massless radiation on the light cone. On the other hand, because of \eqref{IBPIdentity}, at the level of the Born approximation, we see that the geometric tensors contribute only to the tail piece of the Green's function.

\section{Schwarzschild and Kerr Geometries}

As a concrete application of our formalism, in this section we will calculate the null cone and tail pieces of the Green's functions in the weak field limit of the Kerr geometry, to first order in the black hole's mass $M$ and angular momentum $S$. Setting $S$ to zero would then give us the first order in mass result for the weak field Schwarzschild geometry. These results, when pushed to higher orders in $M$ and $S$, would provide us with concrete expressions for the Green's functions to investigate the tail induced self force and more generally, the gravitational $n$-body problem, in the weak field limit background of astrophysical black holes. Strictly speaking, because $S \leq M^2$, a consistent answer for the Green's functions would require at least a second order in $M$ calculation, but since this constitutes a significant computational effort, we shall leave it for a future pursuit.

{\bf Schwarzschild} \quad We begin with a discussion of the Schwarzschild case. If we choose to write the Schwarzschild black hole metric in (Cartesian) isotropic coordinates $(t,\vec{x})$, so that there are no off diagonal terms, we may express
\begin{align}
g_{\mu\nu} = \eta_{\mu\nu} + h_{\mu\nu}
\end{align}
where
{\allowdisplaybreaks\begin{align}
h_{00}  &\equiv \left( \frac{1-\frac{M}{2r}}{1+\frac{M}{2r}} \right)^2-1 \\
        &= -4\frac{M}{2r} + 8 \left(\frac{M}{2r}\right)^2 - 12 \left(\frac{M}{2r}\right)^3 + \dots , \nnn
h_{ij}  &\equiv \eta_{ij} \left( \left( 1+\frac{M}{2r} \right)^4-1 \right) \\
        &= \eta_{ij} \left( 4\frac{M}{2r} + 6 \left(\frac{M}{2r}\right)^2 + 4 \left(\frac{M}{2r}\right)^3 + \dots \right) , \nnn
h_{0i}  &= 0 .
\end{align}}
Here $r \equiv \sqrt{\delta_{ij} x^i x^j}$, $M$ is the mass of the black hole, and we have set Newton's constant to unity, $G_\text{N} = 1$. The power series expansion of $h_{00}$ and $h_{ij}$ can be substituted into the $\widehat{\mathcal{I}}$-integral in \eqref{MasterMatrix_II}. At order $(M/2r)^2$ and beyond, the solution would of course receive contributions from more iterations and high order $h$ terms from the integral equations \eqref{GreensFunctionIntegralEquation_Scalar}, \eqref{GreensFunctionIntegralEquation_Photon}, and \eqref{GreensFunctionIntegralEquation_Graviton}, and would likely involve two or more overlapping $\widehat{\mathcal{I}}$-type integrals. Here we will focus on the first Born approximation.

Within the one scattering approximation, the main technical hurdle to overcome is therefore the class of integrals
\begin{align}
\label{Integrals_1/r^n}
\widehat{\mathbb{I}}_{(n)} \equiv \frac{1}{4\pi} \int_{-1}^{+1} \int_0^{2\pi} \frac{\dd (\cos\theta'') \dd \phi''
}{|\vec{x}''[\rho,\theta'',\phi''] - \vec{z}[s,\theta_+,\phi_+]|^n}
\end{align}
where $n$ is a positive integer, $\vec{x}''$ has Cartesian components defined in \eqref{x''} (so that, in particular, $\rho = \Delta^0 = t-t'$), and
\begin{align}
\vec{z} \equiv -\frac{\vec{x}+\vec{x}'}{2} .
\end{align}
Because we choose our coordinate system such that $\vec{x}-\vec{x}' = |\vec{\Delta}| \widehat{e}_3$, where $\widehat{e}_3$ is the unit vector in the $3$-direction, we have the following equalities (see \eqref{x'''}),
{\allowdisplaybreaks\begin{align}
\vec{z}[s,\theta_+,\phi_+]
&= \frac{1}{2}\bigg(
    \sqrt{s^2-|\vec{\Delta}|^2} \sin\theta_+ \cos\phi_+, \nonumber\\
    &\qquad \sqrt{s^2-|\vec{\Delta}|^2}\sin\theta_+ \sin\phi_+, \nonumber\\
    &\qquad s \cos\theta_+ \bigg) = \frac{|\vec{\Delta}|}{2} \widehat{e}_3 - \vec{x},
\end{align}}
from which we can deduce that
\begin{align}
\label{s=r+r'}
s = r + r', \qquad \cos\theta_+ = \frac{r'-r}{|\vec{x}-\vec{x}'|}.
\end{align}
(The other solution $(s,\cos\theta_+) = (|r-r'|,-(r+r')/|\vec{x}-\vec{x}'|)$ is inadmissible because $(r+r')/|\vec{x}-\vec{x}'| \geq 1$.) Here, $r \equiv |\vec{x}|$ and $r' \equiv |\vec{x}'|$, and the azimuth angles of $\vec{x}$ and $\vec{x}'$ are both equal to $\phi_+ + \pi$.

For the moment, it helps to think of $\vec{x}''$ and $\vec{z}$ as independent vectors which we have chosen to write their Cartesian components in terms of ellipsoidal coordinates $(\rho,\theta'',\phi'')$ and $(s,\theta_+,\phi_+)$; we will also take $R$ in \eqref{EuclideanMetric_EllipsoidalCoordinates} to be simply a constant, not necessarily equal to $|\vec{\Delta}|$.

The $n=1$ case has been evaluated by both DeWitt and DeWitt \cite{DeWittDeWitt:1964} and Pfenning and Poisson \cite{PfenningPoisson:2000zf} by performing a prolate ellipsoidal harmonics expansion of the inverse Euclidean distance $|\vec{x}''-\vec{z}|^{-1}$. An alternate means of getting the same result is, as already noted by DeWitt and DeWitt, to recognize that $\widehat{\mathbb{I}}_{(1)}$ is the Columb (electric) potential of a charged perfectly conducting ellipsoid defined by $\vec{x}''$.\footnote{The Columb potential at $\vec{z}$ can be obtained by the Green's function type integral $\int \dd^2 x'' \sqrt{g_2''} \Sigma[\vec{x}'']/(4\pi|\vec{x}''-\vec{z}|)$, where $g_2$ is the determinant of the induced metric on the ellipsoidal surface and the surface charge density $\Sigma$ is the normal derivative of the electric potential, $\Sigma = N^i \partial_i \Psi$, evaluated on the said surface. Because $N^i \partial_i$ is a unit normal, one would find that the combination $\dd^2 x'' \sqrt{g''_2} N^i \partial_i \Psi$ is equal to the infinitesimal solid angle $\dd \Omega$ in 3 spatial dimensions, up to overall constant factors.} By definition, the conducting surface is an equipotential one. This implies that the answer to $\widehat{\mathbb{I}}_{(1)}$ has to depend on the $s$-coordinate of $\vec{z}$ only, for that would automatically be a constant on the ellipsoidal surface. For $\vec{z}$ lying away from the ellipsoidal surface, our integral must satisfy Poisson's equation $g^{ij} \nabla_{z^i} \nabla_{z^j} \widehat{\mathbb{I}}_{(1)}[s] = 0$ (with the inverse metric $g^{ij}$ of \eqref{EuclideanMetric_EllipsoidalCoordinates}), which in turn is equivalent to the ordinary differential equation
\begin{align}
0 = (1-\kappa^2) \frac{\dd^2 \widehat{\mathbb{I}}_{(1)}[\kappa]}{\dd \kappa^2} - 2 \kappa \frac{\dd \widehat{\mathbb{I}}_{(1)}[\kappa]}{\dd \kappa},
        \quad \kappa \equiv s/R.
\end{align}
The general solution is a linear combination of a constant and the Legendre function $Q_0[s/R] = (1/2)\ln[((s/R)+1)/((s/R)-1)]$. But the asymptotic boundary condition implied by the integral representation in \eqref{Integrals_1/r^n} is
\begin{align}
\lim_{s \to \infty} \widehat{\mathbb{I}}_{(1)} \to \lim_{s \to \infty} \frac{1}{4\pi|\vec{z}|} \int_{\mathbb{S}^2}\dd \Omega \to \frac{2}{s} .
\end{align}
(When $s \gg R$, \eqref{x'''} says $s/2 \approx |\vec{z}|$; $s/2$ essentially becomes the spherical radial coordinate.) The asymptotic limit
\begin{align}
\lim_{s \to \infty} Q_0[s/R] \to \frac{R}{s}
\end{align}
then tells us the solution for $\vec{z}$ located outside the ellipsoid is
\begin{align}
\label{sgreaterthanrho}
\frac{1}{4\pi} \int_{-1}^{+1} \int_0^{2\pi} \frac{\dd (\cos\theta'') \dd \phi''}{|\vec{x}''[\rho,\theta'',\phi''] - \vec{z}[s > \rho,\theta_+,\phi_+]|} \nonumber\\
= \widehat{\mathbb{I}}_{(1)}[s > \rho] = \frac{1}{R} \ln\left[ \frac{s+R}{s-R} \right] .
\end{align}
A conducting surface forms a Faraday cage, so for $\vec{z}$ lying inside the ellipsoid, the potential is position independent and the same as that on the surface,
\begin{align}
\label{slessthanrho}
\frac{1}{4\pi} \int_{-1}^{+1} \int_0^{2\pi} \frac{\dd (\cos\theta'') \dd \phi''}{|\vec{x}''[\rho,\theta'',\phi''] - \vec{z}[s \leq \rho,\theta_+,\phi_+]|} \nonumber\\
= \widehat{\mathbb{I}}_{(1)}[s \leq \rho] = \frac{1}{R} \ln\left[ \frac{\rho+R}{\rho-R} \right] .
\end{align}
Reinstating the relationships $\vec{z} = -(\vec{x}+\vec{x}')/2$, $R = |\vec{x}-\vec{x}'|$, $\rho = t-t'$ and $s = r+r'$, we gather
{\allowdisplaybreaks\begin{align}
\label{I_(1)}
\widehat{\mathbb{I}}_{(1)}
&= |\vec{x}-\vec{x}'|^{-1} \\
&\times \bigg(
\Theta[r+r'-(t-t')] \ln\left[ \frac{r+r'+|\vec{x}-\vec{x}'|}{r+r'-|\vec{x}-\vec{x}'|} \right] \nonumber\\
&\qquad + \Theta[t-t'-(r+r')] \ln\left[ \frac{t-t'+|\vec{x}-\vec{x}'|}{t-t'-|\vec{x}-\vec{x}'|} \right]
\bigg). \nonumber
\end{align}}
For later use, let us record the following symmetrized spatial derivative on $\widehat{\mathbb{I}}_{(1)}$, keeping in mind that $(\partial_{i} + \partial_{i'})$ acting on any function that depends on the spatial coordinates solely through the difference $\vec{x}-\vec{x}'$ is zero:
{\allowdisplaybreaks\begin{align}
\label{I_(1)_Derivative}
(\partial_i + \partial_{i'}) \widehat{\mathbb{I}}_{(1)}
&= |\vec{x}-\vec{x}'|^{-1} \Theta[r+r'-(t-t')] \\
&\qquad \times (\partial_i + \partial_{i'}) \ln\left[ \frac{r+r'+|\vec{x}-\vec{x}'|}{r+r'-|\vec{x}-\vec{x}'|} \right] \nonumber
\end{align}}
(Note that the two $\delta$-function terms arising from differentiating the $\Theta[r+r'-(t-t')]$ and $\Theta[t-t'-(r+r')]$ in $\widehat{\mathbb{I}}_{(1)}$ cancel each other, upon setting $t-t' = r+r'$ in their respective coefficients.)

When $n>1$, this conducting ellipsoid interpretation for the $n=1$ case does not continue to hold; but one may attempt to derive a partial differential equation in terms of the variables $(s,\theta_+,\phi_+)$, such that some differential operator $\mathcal{D}$ acting on the kernel $|\vec{x}''-\vec{z}|^{-n}$ is zero. (Note that if $\vec{x}''$ and $\vec{z}$ lived in $n+2$ spatial dimensions, $\mathcal{D}$ would be the $(n+2)$-dimensional Laplacian, but implementing this scheme would involve introducing an additional $n-1$ fictitious angles and Cartesian components for $\vec{x}''$ and $\vec{z}$.) The general solutions of this partial differential equation may either help lead to a physical interpretation -- just as one was found for the $n=1$ case -- or a harmonics expansion analogous to the one used by DeWitt and DeWitt, so that the resulting series can be integrated term-by-term. Because of the cylindrical symmetry of the integral $\widehat{\mathbb{I}}_{(n)}$, the final result should not depend on $\phi_+$. We shall leave these pursuits for future work, and merely sum up the $\mathcal{O}[M/|\vec{\Delta}|]$ results here. Recalling the relationship between $h_{\alpha\beta}$ and $\widehat{\mathcal{I}}_{\alpha\beta}$ from \eqref{MasterMatrix_II}:
\begin{align}
\widehat{\mathcal{I}}_{\alpha\beta} = -\delta_{\alpha\beta} M \widehat{\mathbb{I}}_{(1)} + \mathcal{O}\left[\left(M/|\vec{x}-\vec{x}'|\right)^2\right],
\end{align}
with $\widehat{\mathbb{I}}_{(1)}$ given by \eqref{I_(1)}.

{\bf Kerr} \quad Let us now turn our attention to a Kerr black hole with mass $M$ and angular momentum $S$, with its spin axis aligned along the $3$-direction.\footnote{This $3$-direction is not to be confused with the $3$-direction of the prolate ellipsoidal coordinate system invoked during the evaluation of $\widehat{\mathcal{I}}_{\alpha\beta}$ in \eqref{MasterMatrix_II}.} Starting from the Kerr metric written in Boyer-Lindquist coordinates (see equation 33.2 of \cite{MisnerThorneWheeler:1974qy}), we first perform the following transformation on the $r$ coordinate
\begin{align}
r \to r \left( 1 + \frac{M}{2r} \right)^2.
\end{align}
(This coordinate transformation would yield, when $S=0$, the Schwarzschild metric in isotropic coordinates.) Denoting the unit vector in the $3$-direction as $\widehat{e}_3$ and further define
\begin{align}
\vec{S} \equiv S \widehat{e}_3,
\end{align}
to first order in both $S$ and $M$, we may then write the Kerr metric as
\begin{align}
\label{WeakKerrMetric_I}
g_{\mu\nu} &= \eta_{\mu\nu} + h_{\mu\nu}
\end{align}
where
\begin{align}
\label{WeakKerrMetric_II}
h_{\alpha\beta}[t,\vec{x}] \approx -2\left( M \delta_{\alpha\beta}
    + \delta^0_{\{\alpha} \delta^i_{\beta\}} \left( \vec{S} \times \frac{\partial}{\partial \vec{x}} \right)_i \right) \frac{1}{r}.
\end{align}
In a Cartesian basis,
\begin{align}
\left(\vec{S} \times \frac{\partial}{\partial \vec{x}}\right)_i = S \left(-\frac{\partial}{\partial x^2},\frac{\partial}{\partial x^1},0 \right)_i.
\end{align}
The off diagonal nature of $\delta^0_{\{\alpha} \delta^i_{\beta\}}$ implies that the first order in mass $\widehat{\mathcal{I}}_{00}$ and $\widehat{\mathcal{I}}_{ij}$ for the Kerr black hole are identical to that of the Schwarzschild case. As for $\widehat{\mathcal{I}}_{0i}$, by referring to \eqref{MasterIntegral}, integrating by parts the spatial gradient acting on $r^{-1}$, and using \eqref{Minkowski_TranslationSymmetry} to pull the resulting two derivatives out of the integral, we observe that it can be gotten by acting
\begin{align}
\mathcal{J}_i \equiv \left\{ \vec{S} \times \left(\frac{\partial}{\partial \vec{x}} + \frac{\partial}{\partial \vec{x}'} \right) \right\}_i
\end{align}
on the $n=1$ integral in \eqref{Integrals_1/r^n}. That is,
\begin{align}
\widehat{\mathcal{I}}_{0i} = -\mathcal{J}_i \widehat{\mathbb{I}}_{(1)} + \dots
\end{align}
Altogether, to first order in mass $M$ and angular momentum $S$, the Kerr spacetime hands us
\begin{align}
\label{Ihat_Kerr}
\widehat{\mathcal{I}}_{\alpha\beta}
    = -\left( M \delta_{\alpha\beta} + \delta_{\{\alpha}^0 \delta_{\beta\}}^i \mathcal{J}_i \right) \widehat{\mathbb{I}}_{(1)},
\end{align}
with $\widehat{\mathbb{I}}_{(1)}$ given by \eqref{I_(1)}.

We will now construct the null cone portion of the Green's functions by computing the world function, van Vleck determinant, and the parallel propagator. From \eqref{WorldFunction_PerturbedMinkowski}, \eqref{vanVleckDeterminant_PerturbedMinkowski} and \eqref{ParallelPropagator_PerturbedMinkowski_I0Form}, we recall that these objects may be gotten once $\widehat{\mathcal{I}}_{\alpha\beta}^{(0)}$ is known. $\widehat{\mathcal{I}}_{\alpha\beta}^{(0)}$ is related to $\widehat{\mathcal{I}}_{\alpha\beta}$, as can be inferred from \eqref{MasterMatrix_II}, by replacing the $|\vec{x}-\vec{x}'|$ in the time argument of $h_{\alpha''\beta''}$ with $t-t'$; and replacing the $t-t'$ in the spatial arguments of $h_{\alpha''\beta''}$ with $|\vec{x}-\vec{x}'|$. Since the $h_{\alpha''\beta''}$ at hand does not have any time dependence, this means $\widehat{\mathcal{I}}_{\alpha\beta}^{(0)}$ is given by replacing every $t-t'$ with $|\vec{x}-\vec{x}'|$ in \eqref{Ihat_Kerr}. Because $|\vec{x}-\vec{x}'| \leq r+r'$, this means the $\Theta[t-t'-(r+r')]$ term in \eqref{I_(1)} may be dropped and the $\Theta[r+r'-(t-t')]$ set to unity.\footnote{The following remark is in order. Because $|r-r'| \leq |\vec{x}-\vec{x}'| \leq r+r'$, the only way $r+r' = t-t' = |\vec{x}-\vec{x}'|$ can be satisfied simultaneously is when a null signal is sent from $\vec{x}'$ to $\vec{x}$ with the spatial origin (i.e. the spatial location of the black hole) lying on the straight line joining them (as viewed in Euclidean $3-$space), so that $\vec{x}\cdot\vec{x}' = -rr'$. But we do not expect any signal to be able to pass through the black hole; hence, all terms implying such a configuration may be discarded.}
\begin{align}
\label{Ihat0_Kerr}
\widehat{\mathcal{I}}_{\alpha\beta}^{(0)}
    &= - \frac{1}{|\vec{x}-\vec{x}'|} \left( M \delta_{\alpha\beta} + \delta_{\{\alpha}^0 \delta_{\beta\}}^i \mathcal{J}_i \right) \nonumber\\
    &\qquad \qquad \times \ln\left[ \frac{r+r'+|\vec{x}-\vec{x}'|}{r+r'-|\vec{x}-\vec{x}'|} \right]
\end{align}

\emph{World Function} \quad The world function is $\sigma_{x,x'} \approx \bar{\sigma}_{x,x'} + \Delta^\alpha \Delta^\beta \widehat{\mathcal{I}}_{\alpha\beta}^{(0)}$. Some calculus reveals
{\allowdisplaybreaks\begin{align}
\label{WorldFunction_Kerr}
&\sigma_{x,x'}
\approx \bar{\sigma}_{x,x'}
- \left( \frac{1}{r} + \frac{1}{r'} \right) \frac{2 (t-t') \vec{S} \cdot \left(\vec{x} \times \vec{x}'\right)}{r r'\left(1 + \widehat{x}\cdot\widehat{x}' \right)} \\
&- \frac{M}{|\vec{x}-\vec{x}'|}  \left( (t-t')^2 + (\vec{x}-\vec{x}')^2 \right)
    \ln\left[ \frac{r+r'+|\vec{x}-\vec{x}'|}{r+r'-|\vec{x}-\vec{x}'|} \right] \nonumber,
\end{align}}
with $\vec{S} \cdot \left(\vec{x} \times \vec{x}'\right) = S(x^1 x'^2 - x'^1 x^2)$, $\widehat{x} \equiv \vec{x}/r$, $\widehat{x}' \equiv \vec{x}'/r'$ and $\widehat{x} \cdot \widehat{x}' \equiv \delta_{ij} \widehat{x}^i \widehat{x}'^j$ being the Euclidean dot product. 

\emph{van Vleck Determinant} \quad Because the Kerr spacetime is a vacuum solution to Einstein's equations $R_{\mu\nu} = 0$, the Ricci tensor to first order in mass and angular momentum must vanish, at least away from the spatial origin $\vec{x} \neq \vec{0}$. Visser's result \eqref{vanVleckDeterminant_PerturbedMinkowski_RicciForm} then informs us that the van Vleck determinant must remain unity to this order,
\begin{align}
\label{vanVleck_Kerr}
\Delta_{x,x'} \approx 1.
\end{align}
We may also confirm this by computing the van Vleck determinant from the world function in \eqref{WorldFunction_Kerr} using \eqref{vanVleckDeterminant_Def}, or by a direct differentiation (see \eqref{vanVleckDeterminant_PerturbedMinkowski})
\begin{align}
&\left( \eta^{\alpha\beta} + \Delta^\beta \left( \partial^\alpha - \partial^{\alpha'} \right)
    -\frac{1}{2} \Delta^\alpha \Delta^\beta \partial^\mu \partial_{\mu'} \right) \widehat{\mathcal{I}}^{(0)}_{\alpha\beta} \nonumber\\
    &\qquad\qquad = \frac{1}{4} (h[x]+h[x']) = M\left( \frac{1}{r}+\frac{1}{r'} \right).
\end{align}

\emph{Parallel Propagator} \quad According to \eqref{ParallelPropagator_PerturbedMinkowski}, the symmetric portion of the parallel propagator can be read off the metric perturbations, namely
{\allowdisplaybreaks\begin{align}
\label{ParallelPropagator_Symmetric_Kerr}
&\frac{1}{2}\left( g_{\mu\nu'} + g_{\nu\mu'} \right) \\
&\qquad = \eta_{\mu\nu}
- M\left(\frac{1}{r}+\frac{1}{r'}\right)\delta_{\mu\nu} \nonumber\\
&\qquad \qquad + \delta^0_{\{\mu}\delta_{\nu\}i}
    \left(\frac{1}{r^3}\left(\vec{S} \times \vec{x}\right)^i + \frac{1}{r'^3}\left(\vec{S} \times \vec{x}'\right)^i\right) \nonumber
\end{align}}
At this point, it is convenient to define
\begin{align}
\label{Vj_Def}
\mathcal{V}_j \equiv \frac{\widehat{x}_j + \widehat{x}'_j}{r r'\left( 1 + \widehat{x} \cdot \widehat{x}' \right)}.
\end{align}
(One may need to recognize $(r+r')-|\vec{x}-\vec{x}'|^2 = 2rr'(1 + \widehat{x} \cdot \widehat{x}')$.) By a direct calculation, one may show that $\mathcal{V}_j$ is divergence-less.
\begin{align}
\label{VisDivergenceless}
\partial^i \mathcal{V}_i = \partial^{i'} \mathcal{V}_i = 0
\end{align}
and it also satisfies
\begin{align}
\label{VCurlD=0}
\vec{S} \cdot \left( \frac{\partial}{\partial \vec{x}} \times \frac{\partial}{\partial \vec{x}'} \right) \mathcal{V}_j = 0.
\end{align}
The antisymmetric portion of the parallel propagator is given by
\begin{align}
\frac{1}{2}\left( g_{\mu\nu'} - g_{\nu\mu'} \right) &= \Delta^\alpha \partial_{[\mu^+} \widehat{\mathcal{I}}_{\nu] \alpha}^{(0)}.
\end{align}
In terms of $\mathcal{V}_i$, its non-zero components are
\begin{align}
\label{ParallelPropagator_AntiSymmetric_I_Kerr}
\frac{1}{2}\left( g_{0j'} - g_{j0'} \right) = \left( M (t-t') + \Delta^i \mathcal{J}_i \right) \mathcal{V}_j
\end{align}
and
\begin{align}
\label{ParallelPropagator_AntiSymmetric_II_Kerr}
\frac{1}{2}\left( g_{jk'} - g_{kj'} \right)
= \left( M \Delta_{[k} - (t-t') \mathcal{J}_{[k} \right) \mathcal{V}_{j]}.
\end{align}

\emph{Tails in Kerr} \quad By recalling \eqref{I_(1)_Derivative}, $\widehat{\mathcal{I}}_{0i} = -\mathcal{J}_i \mathbb{I}_{(1)}$ reads
\begin{align}
\label{KerrIab}
\widehat{\mathcal{I}}_{0i} &= -|\vec{x}-\vec{x}'|^{-1} \Theta[r+r'-(t-t')] \nonumber\\
&\qquad \qquad \times \mathcal{J}_i
\ln\left[ \frac{r+r'+|\vec{x}-\vec{x}'|}{r+r'-|\vec{x}-\vec{x}'|} \right] \nonumber\\
&\quad + \mathcal{O}\left[ \left(S/|\vec{x}-\vec{x}'|^2\right)^2 \right]
\end{align}
To first order in angular momentum $S$, therefore, the Kerr spacetime does add non-trivial terms to the null cone portion of the Green's functions of massless fields in a Schwarzschild spacetime. However, the tail part of these Green's functions only receives additional contributions from the Kerr spacetime within the region $r + r' \geq t-t' > |\vec{x}-\vec{x}'|$ near the null cone; no contributions due to angular momentum arise deeper inside the null cone, $t-t' > r + r'$. This latter observation is consistent with Poisson's findings in \cite{Poisson:2002jz}. In fact, we shall find that the tail of the scalar and photon Green's functions are only altered by angular momentum precisely at $t-t' = r+r'$, corresponding to the reflection of null rays off the black hole. Only the tail of the graviton Green's function, which is sensitive not only to the Ricci curvature but to Riemann as well, experience angular momentum effects throughout the region $|\vec{x}-\vec{x}'| < t-t' \leq r+r'$.

Let us now proceed to compute the various pieces of the tail portion of the Green's functions. Equations \eqref{GreensFunctionBornApproximation_Scalar_eta_Decomposed}, \eqref{GreensFunctionBornApproximation_Photon_eta_Decomposed}, and \eqref{GreensFunctionBornApproximation_Graviton_eta_Decomposed} tell us there are only three distinct building blocks. Employing \eqref{Ihat_Kerr}, these are as follows. The first is
\begin{align}
&\widehat{\mathcal{I}}^{(\text{S})} \equiv \frac{1}{2} \partial^\mu \partial_{\mu'} \widehat{\mathcal{I}} - \partial_\rho \partial_{\kappa'} \widehat{\mathcal{I}}^{\rho\kappa} \nonumber\\
\label{Tail_Scalar}
&= \frac{1}{2}\left( \partial_t - \partial_{t'} \right) \left( \frac{4 M \Theta[t-t'-(r+r')]}{(t-t')^2-|\vec{x}-\vec{x}'|^2} \right) \\
&\qquad + \frac{2 \vec{S} \cdot \left( \widehat{x} \times \widehat{x}' \right)}{r r' \left( 1 + \widehat{x} \cdot \widehat{x}' \right)} \delta'[t-t'-(r+r')], \nonumber
\end{align}

The second is
\begin{align}
\label{Tail_ParallelPropagator_Def}
\widehat{\mathcal{I}}^{(\text{A})}_{\mu\nu} \equiv \frac{1}{2} \partial^{\alpha^-} \partial_{[\mu^+} \widehat{\mathcal{I}}_{\nu]\alpha},
\end{align}
where $\partial_{\mu^\pm} \equiv \partial_\mu \pm \partial_{\mu'}$. In terms of $\mathcal{V}_j$ in \eqref{Vj_Def}, the non-zero components of $\widehat{\mathcal{I}}^{(\text{A})}_{\mu\nu}$ are then
\begin{align}
\label{Tail_ParallelPropagator_0j}
\widehat{\mathcal{I}}^{(\text{A})}_{0j}
&= - M \delta[r+r'-(t-t')] \mathcal{V}_j, 
\end{align}
where \eqref{VCurlD=0} was used to set the angular momentum terms to zero, and
\begin{align}
\label{Tail_ParallelPropagator_jk}
\widehat{\mathcal{I}}^{(\text{A})}_{jk}
&= \frac{M}{2} \partial_{[k^-} \left( \Theta[r+r'-(t-t')] \mathcal{V}_{j]} \right) \\
&\qquad + \mathcal{J}_{[k} \left( \delta[r+r'-(t-t')] \mathcal{V}_{j]} \right). \nonumber
\end{align}
The third and final building blocks are the geometric curvature terms. The non-zero components of the Riemann terms are
{\allowdisplaybreaks\begin{align}
\label{Tail_Riemann0i0j}
\widehat{(R|1)}_{0i0j}
&= \frac{M}{4} \partial_{\{i^+} \left( \Theta[r+r'-(t-t')] \mathcal{V}_{j\}} \right), \\
\label{Tail_Riemann0ijk}
\widehat{(R|1)}_{0ijk}
&= -\frac{1}{2} \partial_{[j^+} \mathcal{J}_{k]} \left( \Theta[r+r'-(t-t')] \mathcal{V}_i \right), \\
\label{Tail_Riemannijkl}
\widehat{(R|1)}_{ijkl}
&= -\frac{M}{2} \partial_{[k^+} \left( \delta_{l]i} \Theta[r+r'-(t-t')] \mathcal{V}_j \right) \nonumber\\
&\qquad \qquad - (i \leftrightarrow j)
\end{align}}
where the $(i \leftrightarrow j)$ means one has to take the preceding term and swap the indices $i$ and $j$. Performing the appropriate contractions and utilizing \eqref{VisDivergenceless} yields the Ricci tensor and scalar terms
{\allowdisplaybreaks\begin{align}
\label{Tail_RicciTensor}
\widehat{(R|1)}_{\alpha\beta} &= \left( M \delta_{\alpha\beta} + \delta^0_{\{\alpha}\delta^i_{\beta\}} S \mathcal{J}_i \right) \frac{\delta[t-t'-(r+r')]}{rr'} \\
\label{Tail_RicciScalar}
\widehat{(\mathcal{R}|1)} &= -2 M \frac{\delta[t-t'-(r+r')]}{rr'}
\end{align}}
We note that these $\delta$-functions (the $\delta[t-t'-(r+r')]$ and its derivative) arise from null rays scattering off the point mass (i.e., the black hole) at the spatial origin. For instance, one may also arrive at \eqref{Tail_RicciTensor} by recalling from \eqref{MasterMatrix_Ricci_I} and \eqref{MasterMatrix_Ricci_II} that the $\widehat{(R|1)}_{\alpha\beta}$ is an integral involving the Ricci tensor over a prolate ellipsoid centered at $(\vec{x}+\vec{x}')/2$ and whose foci are at $\vec{x}$ and $\vec{x}'$. Since the linearized Ricci tensor for the metric in \eqref{WeakKerrMetric_I} and \eqref{WeakKerrMetric_II} is
\begin{align*}
(R|1)_{\alpha\beta}[\vec{x}] = 4\pi \left( M \delta_{\alpha\beta} + \delta^0_{\{\alpha}\delta^i_{\beta\}} \left(\vec{S} \times \frac{\partial}{\partial \vec{x}}\right)_i \right) \delta^{(3)}[\vec{x}]
\end{align*}
(the $4\pi \delta^{(3)}[\vec{x}]$ comes from $-\delta^{ij} \partial_i \partial_j r^{-1}$) we have
\begin{align*}
\widehat{(R|1)}_{\alpha\beta}
&= 4\pi \left( M \delta_{\alpha\beta} + \delta^0_{\{\alpha}\delta^i_{\beta\}} \mathcal{J}_i \right) \nonumber\\
&\qquad \qquad \times \frac{1}{2} \int \frac{\dd \Omega''}{4\pi}  \delta^{(3)}\left[\frac{\vec{x}+\vec{x}'}{2} + \vec{x}'' \right],
\end{align*}
with the $\vec{x}''$ in \eqref{x''}. This integral leads us to \eqref{Tail_RicciTensor}, if we re-express $\delta^{(3)}[\vec{x}''-\vec{z}] = \delta[s-\rho]\delta[\cos\theta''-\cos\theta_+]\delta[\phi_++\pi-\phi''](\sqrt{\det g_{ij}}/\sin\theta'')^{-1}$, using \eqref{EuclideanMetric_EllipsoidalCoordinates_Vol}.

\emph{Green's Functions} \quad We may now put together the minimally coupled massless scalar Green's function in a weak field Kerr spacetime, with the geometry described in \eqref{WeakKerrMetric_I} and \eqref{WeakKerrMetric_II}, to first order its mass $M$ and angular momentum $S$.
\begin{align}
\label{GreensFunction_Kerr_Scalar}
G_{x,x'} &\approx \frac{\Theta[t-t']}{4\pi} \bigg\{ \delta\left[ \sigma_{x,x'} \right] + \Theta\left[\sigma_{x,x'}\right] \widehat{\mathcal{I}}^\text{(S)} \bigg\}
\end{align}
The Lorenz gauge photon counterpart is
\begin{align}
\label{GreensFunction_Kerr_Photon}
&G_{\mu\nu'} \approx \frac{\Theta[t-t']}{4\pi} \bigg\{
g_{\mu\nu'} \delta\left[\sigma_{x,x'}\right] \\
&\qquad \qquad + \Theta\left[\sigma_{x,x'}\right]
    \left( \eta_{\mu\nu} \widehat{\mathcal{I}}^\text{(S)} + \widehat{\mathcal{I}}^\text{(A)}_{\mu\nu} + \widehat{(R|1)}_{\mu\nu} \right)
\bigg\}, \nonumber
\end{align}
while the de Donder gauge graviton's is
\begin{widetext}
\begin{align}
\label{GreensFunction_Kerr_Graviton}
&G_{\delta\epsilon \rho'\sigma'}
\approx \frac{\Theta[t-t']}{4\pi} \bigg\{ P_{\delta\epsilon \rho'\sigma'} \delta\left[ \sigma_{x,x'} \right]
+ \Theta\left[ \sigma_{x,x'} \right] \bigg(
\bar{P}_{\delta\epsilon \rho\sigma} \widehat{\mathcal{I}}^\text{(S)}
+ \frac{1}{2} \bigg(
    \widehat{\mathcal{I}}^\text{(A)}_{\epsilon\rho} \eta_{\sigma\delta}
    + \widehat{\mathcal{I}}^\text{(A)}_{\epsilon\sigma} \eta_{\rho\delta}
    + \widehat{\mathcal{I}}^\text{(A)}_{\delta\rho} \eta_{\sigma\epsilon}
    + \widehat{\mathcal{I}}^\text{(A)}_{\delta\sigma} \eta_{\rho\epsilon} \bigg) \nonumber\\
&\qquad
    + \bar{P}_{\delta\epsilon \rho\sigma} \widehat{(\mathcal{R}|1)}
+ \eta_{\epsilon\delta} \widehat{(R|1)}_{\rho''\sigma''} + \eta_{\rho\sigma} \widehat{(R|1)}_{\delta''\epsilon''}
    - \frac{1}{2} \eta_{\rho\{\delta} \widehat{(R|1)}_{\epsilon''\}\sigma''} - \frac{1}{2} \eta_{\sigma\{\delta} \widehat{(R|1)}_{\epsilon''\}\rho''}
+ \widehat{(R|1)}_{\rho''\{\delta''\epsilon''\}\sigma''} \bigg)\bigg\} . \nonumber
\end{align}
\end{widetext}
We remind the reader that the van Vleck determinant is unity; whereas the world function $\sigma_{x,x'}$ can be found in \eqref{WorldFunction_Kerr}, the parallel propagator $g_{\mu\nu'}$ components in \eqref{ParallelPropagator_Symmetric_Kerr}, \eqref{ParallelPropagator_AntiSymmetric_I_Kerr} and \eqref{ParallelPropagator_AntiSymmetric_II_Kerr}, $\widehat{\mathcal{I}}^{(\text{S})}$ in \eqref{Tail_Scalar}; the components of $\widehat{\mathcal{I}}^{(\text{A})}_{\mu\nu}$ in \eqref{Tail_ParallelPropagator_0j} and \eqref{Tail_ParallelPropagator_jk}; and the components of the geometric terms such as $\widehat{(R|1)}_{\alpha\beta\mu\nu}$ in \eqref{Tail_Riemann0i0j} through \eqref{Tail_RicciScalar}. As a consistency check of the building blocks $\widehat{\mathcal{I}}^\text{(S)}$, $\widehat{\mathcal{I}}^\text{(A)}_{\mu\nu}$ and $\widehat{(R|1)}_{\mu\nu}$, by employing the identities,
\begin{align}
\delta[-z] = \delta[z], \
\delta'[-z] = -\delta'[z], \
\delta''[-z] = \delta''[z],
\end{align}
we have verified that the tail part of our massless scalar and photon Green's functions satisfy \eqref{DivergenceOfPhotonG}, or equivalently,
\begin{align}
\partial_{\nu^+} \widehat{\mathcal{I}}^\text{(S)} + \partial^\mu \left( \widehat{\mathcal{I}}^\text{(A)}_{\mu\nu} + \widehat{(R|1)}_{\mu\nu} \right) = 0.
\end{align}
\begin{figure}
\begin{center}
\includegraphics[width=3in]{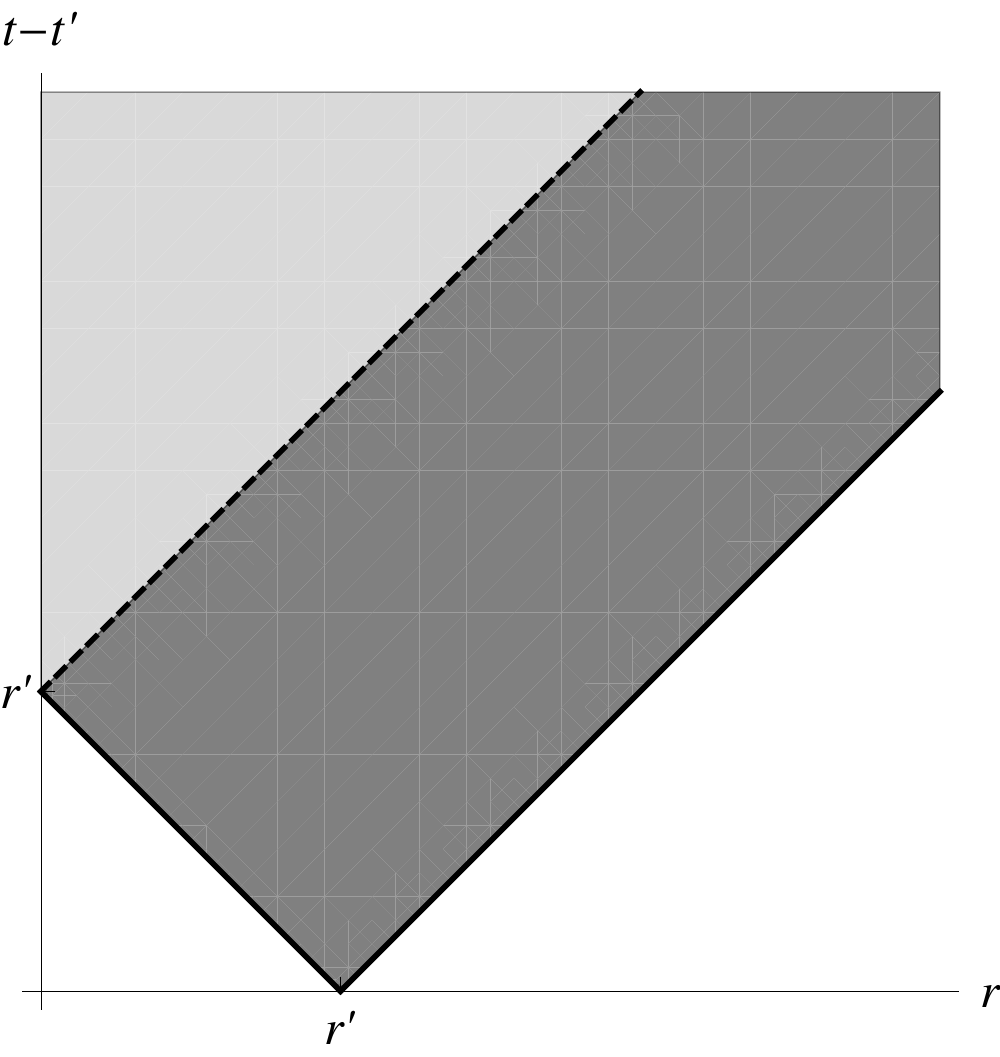}
\end{center}
\caption{Time elapsed ($t-t'$) vs. radial distance ($r$) view of the causal structure of the Green's functions in the weakly curved limit of Kerr spacetime. The spacetime point source is located at radial coordinate distance $r'$ from the black hole. The dark grey area represents the early time tail $|r-r'| \leq |\vec{x}-\vec{x}'| \leq t-t' \leq r+r'$; and the light grey region is the late time tail $t-t' \geq r+r'$. The dashed line is $t-t' = r+r'$; the two solid black lines are $t-t' = |r-r'|$. As already noted by DeWitt and DeWitt \cite{DeWittDeWitt:1964} and Pfenning and Poisson \cite{PfenningPoisson:2000zf}, the Green's functions undergo an abrupt change in behavior when time elapsed transitions from $t-t' < r+r'$ to $t-t' > r+r'$. The region $t-t' \geq r+r'$ only receives contribution from $\widehat{\mathcal{I}}^\text{(S)}$ in \eqref{Tail_Scalar}, which contains angular momentum $S$ terms only when $t-t'$ is exactly equal to $r+r'$; while the dark grey region $|r-r'| \leq t-t' \leq r+r'$ only receives contribution from $\widehat{\mathcal{I}}^\text{(A)}_{\mu\nu}$ in \eqref{Tail_ParallelPropagator_Def} and the geometric curvature terms in \eqref{Tail_Riemann0i0j} through \eqref{Tail_RicciScalar}. The tail part of the massless scalar Green's functions is only non-zero for $t-t' \geq r+r'$, because it is entirely governed by $\widehat{\mathcal{I}}^\text{(S)}$. Because the photon and graviton Green's functions depend on $\widehat{\mathcal{I}}^\text{(A)}_{\mu\nu}$ and the geometric terms, they are non-zero within $|r-r'| \leq t-t' \leq r+r'$. However, their behaviors are altered once $t-t' \geq r+r'$, since they too become governed solely by $\widehat{\mathcal{I}}^\text{(S)}$.}
\label{TailPartOfKerrGreensFunctions}
\end{figure}
\emph{Causal structure} \quad Notice the $\widehat{\mathcal{I}}^{(\text{S})}$ in \eqref{Tail_Scalar} is zero close to the light cone, and only non-zero for late times: $t-t' \geq r+r'$. This in turns indicates the tail part of the massless scalar Green's function is non-zero only after the time elapsed $t-t'$ equals or exceeds the time needed for a null ray to travel from the source at $x'$, reflect off the black hole, and reach the observer at $x$. Furthermore it is sensitive to first order spin effects only exactly at $t-t' = r+r'$. On the other hand, the photon Green's functions contain, in addition to $\widehat{\mathcal{I}}^{(\text{S})}$, the $\widehat{\mathcal{I}}^{(\text{A})}_{\mu\nu}$ in \eqref{Tail_ParallelPropagator_0j} and \eqref{Tail_ParallelPropagator_jk} and $\widehat{(R|1)}_{\alpha\beta}$ in \eqref{Tail_RicciTensor}; the graviton Green's function contain all three building blocks, $\widehat{\mathcal{I}}^{(\text{S})}$, $\widehat{\mathcal{I}}^{(\text{A})}_{\mu\nu}$ and the curvature terms in \eqref{Tail_Riemann0i0j} through \eqref{Tail_RicciScalar}. The $\widehat{\mathcal{I}}^{(\text{A})}_{\mu\nu}$ and curvature terms are non-zero only at early times $|r-r'| \leq |\vec{x}-\vec{x}'| < t-t' \leq r+r'$; mathematically this is because all these terms contain the derivative $\partial_{j^+} \widehat{\mathbb{I}}_{(1)} \propto \Theta[r+r'-(t-t')] \mathcal{V}_j$. However, it may be worthwhile to search for a more physical explanation, for it could lead us to a deeper understanding of the tail effect. In any case, this means both the photon and graviton Green's functions carry non-zero tails throughout the entire interior of the future null cone of $x'$, though their behaviors are altered abruptly when the time elapsed $t-t'$ changes from $t-t' < r+r'$ to $t-t' > r+r'$. This is because, in the former, they contain effects described by $\widehat{\mathcal{I}}^{(\text{A})}_{\mu\nu}$ and geometric curvature; while in the latter region they are, like their scalar cousin, governed solely by $\widehat{\mathcal{I}}^{(\text{S})}$. (We illustrate this abrupt change in behavior of the Green's functions in Figure \eqref{TailPartOfKerrGreensFunctions}.) Finally, we observe that spin effects are present on the null cone and, in the tail, exactly at $t-t' = r+r'$, for the scalar and photon Green's functions. Only the graviton is sensitive to the full Riemann curvature of spacetime, which unlike the Ricci tensor and scalar, is non-zero everywhere. This is why the tail of the graviton Green's function contain spin effects within the whole region of $|\vec{x}-\vec{x}| < t-t' \leq r+r'$.

\section{Summary and concluding thoughts}

In this paper, we have developed a general Born series expansion for solving the minimally coupled massless scalar, photon, and graviton Green's function in perturbed spacetimes described by the metric $g_{\mu\nu} = \gb_{\mu\nu} + h_{\mu\nu}$. The key starting points are the integral equations for the scalar \eqref{GreensFunctionIntegralEquation_Scalar}, photon \eqref{GreensFunctionIntegralEquation_Photon} and graviton \eqref{GreensFunctionIntegralEquation_Graviton} cases, which were gotten from the quadratic portions of the actions of the respective field theories. From these, one performs a power series in the perturbation $h_{\mu\nu}$ and iterate these equations (followed by dropping the remainder terms) however many times necessary to achieve the desired accuracy. We derived a first order integral representation for the scalar \eqref{GreensFunctionBornApproximation_Scalar_gb} and photon \eqref{GreensFunctionBornApproximation_Photon_gb+h} Green's functions in generic backgrounds, and for scalar \eqref{GreensFunctionBornApproximation_Scalar_eta}, photon \eqref{GreensFunctionBornApproximation_Photon_eta+h} and graviton \eqref{GreensFunctionBornApproximation_Graviton_eta+h} in a Minkowski background. Furthermore, in \eqref{GreensFunctionBornApproximation_Scalar_eta_Decomposed}, \eqref{GreensFunctionBornApproximation_Photon_eta_Decomposed}, and \eqref{GreensFunctionBornApproximation_Graviton_eta_Decomposed}, we decomposed these perturbed Minkowski results into their light cone and tail pieces, showing their consistency with the Hadamard form. We reiterate that, at first order in metric perturbations, the solution of the scalar, photon and graviton Green's functions is reduced to the evaluation of the single matrix integral in \eqref{MasterMatrix_II}; the remaining work is mere differentiation. Even though we have applied our perturbation theory only to massless scalars, photons and gravitons, because all we have exploited are the quadratic actions of the field theories involved, our methods should in fact apply to any field theory whose quadratic action is hermitian.

As a concrete application of our formalism, we have calculated the Green's functions of the massless scalar \eqref{GreensFunction_Kerr_Scalar}, photon \eqref{GreensFunction_Kerr_Photon}, and graviton \eqref{GreensFunction_Kerr_Graviton} in the weak field limit of the Kerr black hole geometry, to first order in its mass $M$ and angular momentum $S$. A subset of these weak field results for the Schwarzschild case have previously been obtained by DeWitt and DeWitt \cite{DeWittDeWitt:1964}, and Pfenning and Poisson \cite{PfenningPoisson:2000zf}. Our Kerr calculation shows that, to first order in angular momentum $S$, there will be rotation-induced corrections to these Schwarzschild Green's functions, only on and near the null cone, namely $|\vec{x}-\vec{x}'| \leq t-t' \leq r+r'$ (where $r \equiv |\vec{x}|$ and $r' \equiv |\vec{x}'|$). Beyond that, $t-t' > r+r'$, the behavior of the Green's functions changes abruptly and is governed solely by the mass of the black hole.

Of the previous approaches we have studied -- DeWitt and DeWitt \cite{DeWittDeWitt:1964}, Kovacs and Thorne \cite{KovacsThorne:1975} and Pfenning and Poisson \cite{PfenningPoisson:2000zf} -- DeWitt and DeWitt's seems to be the most general. They utilized Julian Schwinger's perspective that the Green's function is an operator in a fictitious Hilbert space, for example, $G_{x,x'} = \langle x |\widehat{G}| x' \rangle$, from which they found its variation. However, on the level of classical field theory, the main concern of this paper, our methods do not require any additional structure than the quadratic action of the field theory at hand. Hence, we hope it is accessible to a wider audience.\footnote{At the same time, we should mention that Schwinger's \cite{Schwinger:1960qe} initial value formulation of quantum field theory (nowadays known as the Schwinger-Keldysh formalism), has in fact been employed to tackle the post-Newtonian program in general relativity, itself a weak field, perturbative problem about flat spacetime. See, for instance, Galley and Tiglio \cite{Galley:2009px}. There is very likely a position-space diagrammatic calculation one can do to reproduce \eqref{GreensFunctionBornApproximation_Scalar_eta}, \eqref{GreensFunctionBornApproximation_Photon_eta+h} and \eqref{GreensFunctionBornApproximation_Graviton_eta+h}.} Our null cone versus tail decomposition was modeled after the work of Pfenning and Poisson \cite{PfenningPoisson:2000zf} (except we generalized it to arbitrary metric perturbations), who in turn state that their work was based on calculations by Kovacs and Thorne \cite{KovacsThorne:1975}. In Pfenning and Poisson's work, they wrote down a perturbative version of the differential equations in \eqref{GreensFunctionPDE_Scalar}, \eqref{GreensFunctionPDE_Photon} and \eqref{GreensFunctionPDE_Graviton} for a weakly curved spacetime with only scalar perturbations $\Phi$, and derived integral representations of the solutions using the flat spacetime Green's function $\bG_{x,x'}$; their methods can very likely be generalized to arbitrary perturbations. However, repeated (and un-necessary) use was made of the equations obeyed by the gravitational potential $\Phi$. We feel this obscures the fact that the solution of the Green's function of some field theory depends on the geometry but not on the underlying dynamics of the geometry itself.

\begin{figure}
\includegraphics[width=3.5in]{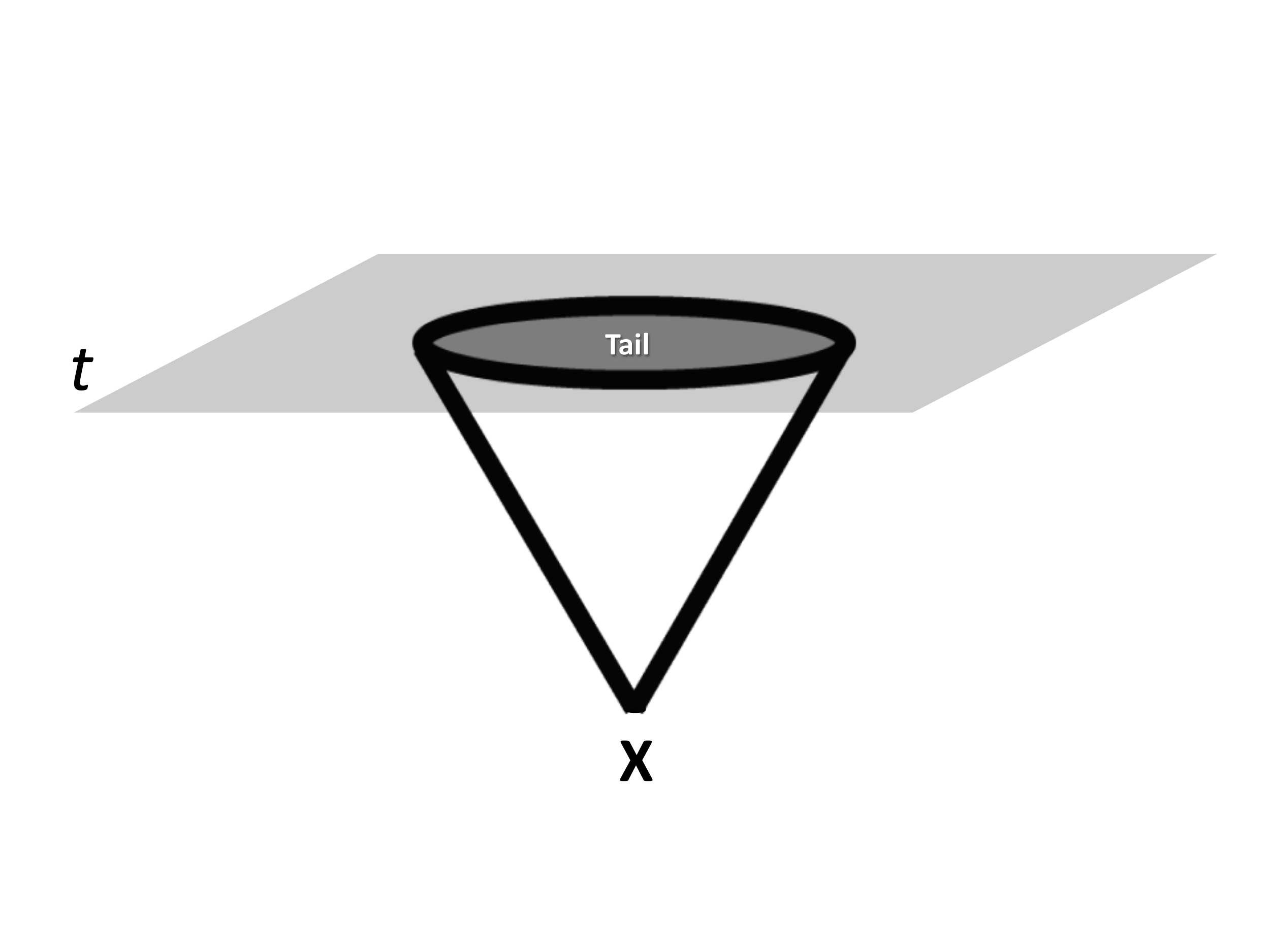}
\caption{At `X', let there be a burst of photons from a source of finite duration. If there were no tails, these photons would sweep out a light cone of non-zero thickness proportional to the duration of the event itself. Because of the metric perturbations $h_{\mu\nu}$, we have shown via our photon Green's function calculation that light develops a tail in a spatially flat inhomogeneous FLRW universe. The dark oval represents the tail of the photon field $A_\mu^\text{(tail)}$ at the present time $t$. Deep within the light cone, we argue that the size of the tail effect in our universe is primarily governed by the power spectrum of the metric perturbations, which is currently being probed by large scale structure observations.}
\label{TailsInCosmologyFigure}
\end{figure}

{\bf Cosmology} \quad We close with some thoughts on applying our work to cosmological physics. We have already shown that the classical theory of light in a spatially flat inhomogeneous FLRW universe is equivalent to that in a perturbed Minkowski spacetime. Consider a source of photons that turns on for a finite duration of time, say a gamma ray burst at redshift $z=6$. We display in Fig. \eqref{TailsInCosmologyFigure} that not only would these photons sweep out a null cone of finite thickness proportional to the duration of the burst, but they will also fill its interior. If $t$ is the present time, the dark oval represents the light that has leaked off the light cone. From our calculation in \eqref{GreensFunctionBornApproximation_Photon_eta_Decomposed}, the tail part of the Green's function and hence the vector potential $A_\mu^\text{(tail)}$ begins at $\mathcal{O}[h]$. Because the components of the stress energy tensor of the electromagnetic fields in an orthonormal frame $T_{\widehat{\mu}\widehat{\nu}}$ (which is what an observer can measure) is quadratic in the derivatives of the potential, $T_{\widehat{\mu}\widehat{\nu}} \sim a^{-4} (\partial A)^2$, this means deep in the interior of the null cone $T_{\widehat{\mu}\widehat{\nu}}$ itself must be quadratic in the metric perturbations $h_{\mu\nu}$.\footnote{A consistent $\mathcal{O}[h^2]$ calculation of the stress energy tensor that is valid everywhere, both near the light cone and deep within it, would therefore require the knowledge of the photon Green's function to $\mathcal{O}[h^2]$.} In cosmology, because the metric perturbations $h_{\mu\nu}$ are believed to be sourced by quantum fluctuations of fields in the very early universe, the $h_{\mu\nu}$ at a particular point in spacetime is a random variable and to obtain concrete results one would have to discuss the statistical average of the product of $h_{\mu\nu}$ with itself, i.e. $\langle h_{\mu\nu}[x'] h_{\alpha\beta}[x''] \rangle$, the power spectrum. The scalar sector of this power spectrum is being probed by the observations currently underway of large scale structure in the universe, and one would have to fold these data into a theoretical investigation of how large the tail effect is in our universe.

One way to proceed is perhaps, following Poisson \cite{Poisson:2002jz}, to start with some generic localized wave packet to mimic the light from a finite duration event such as our gamma ray burst at $z=6$. We may then invoke the Kirchhoff representation in \eqref{KirchoffRep_Photon} (with no current, $J^{\nu'} = 0$) to evolve this wave packet forward in time. At some later time, one can compute the ratio of energies in the tail piece to that still remaining on the null cone
\begin{align}
\left.
\frac{\int T^\text{(tail)}_{\widehat{0}\widehat{0}} \sqrt{-\det[\eta_{ij} + h_{ij}]}\dd^3 x}{\int T^\text{(light cone)}_{\widehat{0}\widehat{0}} \sqrt{-\det[\eta_{ij} + h_{ij}]}\dd^3 x}
\right\vert_t.
\end{align}
This will indicate if there is a significant correction factor that needs to be applied to observations of objects at cosmological distances, when inferring their true brightness. Because of the integrals encountered in \eqref{KirchoffRep_Photon} and the complicated terms in \eqref{GreensFunctionBornApproximation_Photon_eta_Decomposed}, however, this is a difficult calculation. We hope to report on this line of investigation in a future publication.

\section{Acknowledgements}

YZC was supported by funds from the University of Pennsylvania; and by the US Department of Energy (DOE) both at Arizona State University during the 2010-2011 academic year and, before that, at Case Western Reserve University (CWRU), where this work was started. GDS was supported by a grant from the DOE to the particle-astrophysics group at CWRU.

YZC would like to acknowledge discussions with numerous people on the tail effect in cosmology, perturbation theory for Green's functions, and related issues. A non-exhaustive list includes: Niayesh Afshordi, Yi-Fu Cai, Shih-Hung (Holden) Chen, Scott Dodelson, Sourish Dutta, Chad Galley, Ted Jacobson, Justin Khoury, Harsh Mathur, Vincent Moncrief, Eric Poisson, Zain Saleem, and Tanmay Vachaspati.

\appendix

\section{The Matrix $\mathcal{I}_{\alpha\beta}$}
\label{Appendix_TheMatrix}

The primary objective in this section is the analysis of $\mathcal{I}_{\alpha\beta}$ in \eqref{MasterIntegral}, including its behavior near the null cone $\bar{\sigma}_{x,x'} = 0$. A similar discourse may be found in Pfenning and Poisson \cite{PfenningPoisson:2000zf}, but ours is more general because we performed it for arbitrary metric perturbations.

Let us first display the integral in its most explicit form, using the second equality of \eqref{GreensFunctionMinkowski_Scalar}:
\begin{align}
\label{Iab_RawForm}
\mathcal{I}_{\alpha\beta}
    &= \frac{1}{4\pi} \int \dd^4 x'' h_{\alpha''\beta''}[t'',\vec{x}''] \nnn
    &\times \frac{\delta[t-t''-|\vec{x}-\vec{x}''|] \delta[t''-t'-|\vec{x}''-\vec{x}'|]}{|\vec{x}''-\vec{x}'| |\vec{x}-\vec{x}''|} .
\end{align}
We may integrate over $t''$ immediately, so that $t'' = t - |\vec{x}-\vec{x}''| = t' + |\vec{x}''-\vec{x}'|$. This in turns yields the constraint that, viewed in Euclidean 3-space, the observer at $\vec{x}$ and the emitter at $\vec{x}'$ form the foci of a prolate ellipsoid, with semi-major axis $\Delta^0/2$, defined by
\begin{align}
\label{ProlateEllipsoid}
t-t' = |\vec{x}-\vec{x}''| + |\vec{x}''-\vec{x}'| .
\end{align}
This implies, to get a non-zero $\mathcal{I}_{\alpha\beta}$, $x$ needs to lie in the future light cone of $x'$,
\begin{align}
\label{LightConeCondition}
t-t' \geq |\vec{x}-\vec{x}'| .
\end{align}
For by Cauchy's inequality, $|\vec{x}-\vec{x}'| \leq |\vec{x}-\vec{x}''| + |\vec{x}'-\vec{x}''|$, which means, outside the light cone $t-t' < |\vec{x}-\vec{x}'| \leq |\vec{x}-\vec{x}''| + |\vec{x}'-\vec{x}''|$ and no solution can be found. Fig. \eqref{TwoLightConesIntersectFigure} illustrates the situation at hand: we see that the product $\bG_{x,x''} h_{\alpha''\beta''} \bG_{x'',x'}$, due to the causal structure of the Green's function, is non-zero if and only if the $x''$ lie both on the future null cone of $x'$ and on the past null cone of $x$. This can be satisfied if and only if $x'$ lies on or within the past light cone of $x$ or equivalently, if and only if $x$ lies on or within the future light cone of $x'$.

If we now assume that \eqref{LightConeCondition} holds, then it is the surface of the ellipsoid in \eqref{ProlateEllipsoid} that we need to integrate over, weighted by $h_{\alpha''\beta''}[t'',\vec{x}'']$. To see this, let us employ ellipsoidal coordinates centered at $(1/2)(\vec{x} + \vec{x}')$, i.e. put $\vec{x}'' \equiv (1/2)(\vec{x}'+\vec{x}) + \vec{x}'''$, with
\begin{align}
\label{x'''}
\vec{x}'''[s,\theta,\phi]
&= \bigg( \sqrt{(s/2)^2 - (R/2)^2} \sin \theta \cos \phi, \nnn
&\qquad \sqrt{(s/2)^2 - (R/2)^2} \sin \theta \sin \phi, \nnn
&\qquad (s/2) \cos \theta \bigg) .
\end{align}
These coordinates fix the foci to be at $\vec{x}$ and $\vec{x}'$ but allow the size of the ellipse to vary with $s$. (The $1$- and $2$-components of $\vec{x}'''$ tell us $\sqrt{(s/2)^2 - (R/2)^2}$ act as the radial coordinate in the $12$-plane, and hence we shall require $s \geq R$. This means all volume integrals involve $s$ would have limits $\int_{R}^\infty \dd s$.) The Euclidean spatial metric in 3 dimensions goes from $g_{ij} = \delta_{ij}$ for Cartesian coordinates to
{\allowdisplaybreaks\begin{align}
\label{EuclideanMetric_EllipsoidalCoordinates}
g_{ij} &= \text{diag}\bigg[
\frac{(s/2)^2 - (R/2)^2 \cos^2\theta}{s^2-R^2}, \nnn
&\qquad \qquad (s/2)^2 - \left( R/2 \right)^2 \cos^2\theta, \nnn
&\qquad \qquad \left( \left( s/2 \right)^2 - \left( R/2 \right)^2 \right) \sin^2\theta
\bigg], \nnn
&\qquad \quad (s, \theta, \phi), \quad s \geq R ,
\end{align}}
where $R \equiv |\vec{\Delta}| = |\vec{x}-\vec{x}'|$. The Euclidean volume measure is
\begin{align}
\label{EuclideanMetric_EllipsoidalCoordinates_Vol}
\sqrt{\det[g_{ij}]} = \frac{1}{2}
\left( \left( s/2 \right)^2 - \left( R/2 \right)^2 \cos^2\theta \right)\sin\theta .
\end{align}
Using the expressions for the components of $\vec{x}''$ in \eqref{x'''}, we may obtain
\begin{align*}
|\vec{x}''-\vec{x}|     = \frac{s}{2} - \frac{R}{2} \cos[\theta], \quad
|\vec{x}''-\vec{x}'|    = \frac{s}{2} + \frac{R}{2} \cos[\theta] .
\end{align*}
This means the argument in the remaining $\delta$-function of the $\mathcal{I}$-integrand is $t-t'-s$, and the $\sqrt{\det g_{ij}} |\vec{x}-\vec{x}''|^{-1} |\vec{x}'-\vec{x}''|^{-1} = (\sin\theta)/2$. The integral over $s$ can be performed immediately, and because the lower limit is $R$, it gives us $\Theta[t-t'-R] = \Theta[t-t']\Theta[\bsigma_{x,x'}]$ multiplying $h_{\alpha\beta}$ with $s = t-t'$. We are left with the angular integration,
{\allowdisplaybreaks\begin{align}
\label{Ihat_Integral}
&\mathcal{I}_{\alpha\beta}
        \equiv \Theta[t-t']\Theta[\bar{\sigma}_{x,x'}] \widehat{\mathcal{I}}_{\alpha\beta} \nnn
&= \Theta[t-t']\Theta[\bar{\sigma}_{x,x'}] \\
&\times \frac{1}{2} \int_{\mathbb{S}^2} \frac{\dd \Omega}{4\pi}
h_{\alpha''\beta''}
\left[
\frac{t+t'}{2} + \frac{R}{2} \cos[\theta],
\frac{\vec{x} + \vec{x}'}{2} + \vec{x}'''
\right] , \nn
\end{align}}
where now $\vec{x}''' = \vec{x}'''[t-t',\theta,\phi]$.

{\bf Small $\bar{\sigma}_{x,x'}$ expansion} \quad For small $\bar{\sigma}_{x,x'}$, we may develop $\widehat{\mathcal{I}}_{\alpha\beta}$ as a series expansion in powers of $\bar{\sigma}_{x,x'}$. Right on the null cone $t-t' = |\vec{x}-\vec{x}'|$, and if we lie $\vec{x}-\vec{x}'$ along the positive $3$-axis, the spacetime arguments of $h_{\alpha\beta}$ take on the Cartesian components
\begin{align}
\bar{\xi}^\alpha \equiv
    \left( \frac{t+t'}{2} + \frac{\Delta^0}{2} \cos\theta, 0, 0, \frac{x^3+x'^3}{2} + \frac{R}{2} \cos\theta \right)^\alpha
\end{align}
which is equivalent to $\bar{\xi}[\cos\theta] = (1/2)(x+x') + (1/2)(x-x') \cos\theta$. This is a straight line joining $x'$ to $x$. Let us expand about this straight line by expressing the time component $(1/2)(t+t'+R\cos\theta)$ as
\begin{align}
\frac{t+t'}{2} + \sqrt{\left(\frac{t-t'}{2}\right)^2-\frac{\bar{\sigma}_{x,x'}}{2}} \cos\theta
\end{align}
and the $3$-component $(1/2)(x^3+x'^3+\Delta^0 \cos\theta)$ as
\begin{align}
\frac{x^3+x'^3}{2} + \sqrt{\left(\frac{R}{2}\right)^2 + \frac{\bar{\sigma}_{x,x'}}{2}} \cos\theta .
\end{align}
Perform a change of variables in \eqref{Ihat_Integral} $\cos\theta \equiv 2\lambda-1$ and Taylor expand $h_{\alpha\beta}$ in powers of $\bar{\sigma}_{x,x'}$ in the time and $3$-components and in powers of $\sqrt{\bar{\sigma}_{x,x'}}$ in the remaining orthogonal directions. One would find it is necessary to expand the orthogonal directions up to second order to achieve a non-zero result. With $\bar{\xi} = x' + \lambda(x-x')$, the expansion of \eqref{Ihat_Integral} is
{\allowdisplaybreaks\begin{align}
&\widehat{\mathcal{I}}_{\alpha\beta} = \frac{1}{2} \int_0^1 \dd \lambda h_{\alpha''\beta''}\left[\bar{\xi}\right] \\
&+ \frac{\bar{\sigma}_{x,x'}}{4} \int_0^1 \dd \lambda
    \left( \frac{2\lambda-1}{R} \partial_{3''} h_{\alpha''\beta''} - \frac{2\lambda-1}{t-t'} \partial_{0''} h_{\alpha''\beta''} \right) \nonumber\\
&+ \frac{\bar{\sigma}_{x,x'}}{4} \int_0^1\dd \lambda (1-\lambda)\lambda \left( \partial_{1''}^2 + \partial_{2''}^2 \right) h_{\alpha''\beta''} + \dots \nonumber
\end{align}}
where $\partial_{1''}^2 + \partial_{2''}^2$ is the Laplacian involving only the directions orthogonal to $\vec{x}-\vec{x}'$. The arguments of the $h_{\alpha''\beta''}$s on the second and third lines have been suppressed; they are the same as that of the first line. The single derivative terms can be converted into double derivatives by using \eqref{DerivativeLambda} and integrating-by-parts. Up to a remainder that is of $\mathcal{O}[\bar{\sigma}^2]$, one can show
\begin{align}
&\frac{\bar{\sigma}_{x,x'}}{4} \int_0^1 \dd \lambda
    \left( \frac{2\lambda-1}{R} \partial_{Z''} h_{\alpha''\beta''} - \frac{2\lambda-1}{t-t'} \partial_{t''} h_{\alpha''\beta''} \right) \nonumber\\
&\qquad \approx \frac{\bar{\sigma}_{x,x'}}{4} \int_0^1 \dd\lambda (1-\lambda)\lambda \left( \partial_{Z''}^2 - \partial_{t''}^2 \right) h_{\alpha''\beta''} .
\end{align}
By the chain rule,
\begin{align}
(1-\lambda)\lambda \eta^{\mu\nu} \frac{\partial^2 h_{\alpha''\beta''}[\bar{\xi}]}{\partial \bar{\xi}^\mu \partial \bar{\xi}^\nu}
    = \eta^{\mu\nu} \frac{\partial^2 h_{\alpha''\beta''}[\bar{\xi}]}{\partial x^\mu \partial x'^\nu} ,
\end{align}
and hence we gather, as $\bar{\sigma}_{x,x'} \to 0$,
\begin{align}
\widehat{\mathcal{I}}_{\alpha\beta} \approx \left( 1 - \frac{\bar{\sigma}_{x,x'}}{2} \eta^{\mu\nu} \partial_\mu \partial_{\nu'} \right) \left( \frac{1}{2} \int_0^1 h_{\alpha''\beta''}[\bar{\xi}] \dd \lambda \right) .
\end{align}

\end{document}